\documentclass[12pt,preprint]{aastex}

\usepackage{amsmath}
\usepackage{graphicx}
\usepackage{epsfig}
\usepackage{bm}
\usepackage{color}

\newcount\doepsf
\newcommand{\be}{\begin{equation}}
\newcommand{\ee}{\end{equation}}
\newcommand{\bea}{\begin{eqnarray}}
\newcommand{\eea}{\end{eqnarray}}
\newcommand{\bean}{\begin{eqnarray*}}
\newcommand{\eean}{\end{eqnarray*}}

\begin{document}

\title{Heating mechanisms in the low solar atmosphere through magnetic reconnection in current sheets}

\author{Lei Ni$^{1,2}$,
        Jun Lin$^{1}$,
         Ilia I. Roussev$^{3}$,
         and Brigitte Schmieder$^{4}$}

\affil{$^1$Yunnan Observatories, Chinese Academy of Sciences,
                 Kunming 650011, China}
\affil{$^2$Key Laboratory of Solar Activity, National
           Astronomical Observatories, Chinese Academy of Sciences,
           Beijing 100012, China}
\affil{$^3$Division of Geosciences, National Science Foundation
Arlington, Virginia, USA}           
\affil{$^4$Observatoire de Paris, LESIA, Meudon, France}

\shorttitle{The localized heating}  
\shortauthors{Ni et al.}

\email{leini@ynao.ac.cn}

\slugcomment{Localized heating; v.\ \today}

\begin{abstract}
 We simulate several magnetic reconnection processes in the low solar chromosphere/photosphere, the radiation cooling, heat conduction and ambipolar diffusion are all included. Our numerical results indicate that both the high temperature($ \gtrsim 8\times10^4$~K) and low temperature($\sim 10^4$~K) magnetic reconnection events can happen in the low solar atmosphere ($100\sim600$~km above the solar surface). The plasma $\beta$ controlled by plasma density and magnetic fields is one important factor to decide how much the plasma can be heated up. The low temperature event is formed in a high $\beta$ magnetic reconnection process, Joule heating is the main mechanism to heat plasma and the maximum temperature increase is only several thousand Kelvin. The high temperature explosions can be generated in a low $\beta$ magnetic reconnection process, slow and fast-mode shocks attached at the edges of the well developed plasmoids are the main physical mechanisms to heat the plasma from several thousand Kelvin to over $8\times10^4$~K. Gravity in the low chromosphere can strongly hinder the plasmoind instability and the formation of slow-mode shocks in a vertical current sheet. Only small secondary islands are formed; these islands, however, are not well developed as those in the horizontal current sheets. This work can be applied for understanding the heating mechanism in the low solar atmosphere and could possibly be extended to explain the formation of common low temperature EBs ($\sim10^4$~K) and the high tenperature IRIS bombs ($\gtrsim 8\times10^4$) in the future.   
\end{abstract}   


\section{Introduction}
\label{s:introduction}
Magnetic reconnection is an important mechanism to explain many activities and eruptions in the solar atmosphere \citep{2000mrmt.conf.....P}. As the resolution of solar telescopes continue to be improved, many tiny activities relating with magnetic reconnection in the lower solar atmosphere have been observed, {\it e.g.},  chrompshere jets \citep{2009ApJ...707L..37L, 2012A&A...543A...6M, 2013A&A...552L...1B}, Ellerman Bombs (EBs) \citep{2006ApJ...643.1325F, 2007A&A...473..279P, 2014ApJ...792...13H, 2015ApJ...798...19N} and type II white light flares \citep{1994SoPh..149..143D,1999ApJ...512..454D}. Recently, the clear magnetic reconnection events in the low solar atmosphere have been observed by the high resolution Chinese New Vacuum Solar Telescope (NVST), the current sheet structures, inflow and outflow have been clearly identified \citep{2015ApJ...798L..11Y}. Since part of the magnetic energy is converted into plasma's  thermal energy, the local temperature is increased during these magnetic reconnection processes and Joule heating is usually considered to be the main mechanism to heat plasma.

The compact bright points which have UV counterparts and are observed in transition region lines with the Interface Region Imaging Spectrograph (IRIS) satellite are called IRIS bombs \citep[e.g.,][]{2014Sci...346C.315P, 2015ApJ...812...11V, 2016ApJ...824...96T, 2016 A&A...}. While the temperature of the traditional EBs is only around $10^4$~K, the temperature of the IRIS bombs identified in  Si IV slit-jaws  is one magnitude higher \citep[e.g.,][]{2014Sci...346C.315P, 2015ApJ...812...11V, 2016ApJ...824...96T, 2016 A&A...}. \cite{2015ApJ...812...11V} and \cite{2016 A&A...} show that some of the high temperature IRIS bombs ($\gtrsim 8\times10^4$) are possibly due to small flaring arch filaments in the high chromosphere or transition region. However, some of them are still believed to be formed by magnetic reconnection at around the temperature minimum region or even in the photosphere \citep[e.g.,][]{2015ApJ...812...11V, 2016ApJ...824...96T, 2016 A&A...}.

Many jet-like structures are observed in the solar chromosphere \citep[e.g.][]{2012A&A...543A...6M, 2013A&A...552L...1B, 2014Sci...346A.315T}. Compared with corona jets, they are tiny and some with a length even less than $1^{\prime \prime}$ and width less than $0.3^{\prime \prime}$ \citep{2013A&A...552L...1B, 2014Sci...346A.315T}. The life time of these micro jets could be only around one minute \citep{2013A&A...552L...1B, 2014Sci...346A.315T}. The rise speed of these jets is usually in a range $10-100$~km s$^{-1}$ \cite{2012A&A...543A...6M}, and the temperature of the chromosphere material can be heated to higher than $10^5$~K \citep{2012A&A...543A...6M, 2014Sci...346A.315T}. However, magnetic reconnection takes place at which height of the solar chromosphere and where the high temperature plasmas are generated in the chromosphere jets are still not clear. 

Since the density stratification is strong in the chromosphere layer, the plasma density at the top of chromosphere is around 6 magnitudes lower than that at the bottom of the chromosphere \citep{1981ApJS...45..635V, 1993ApJ...406..319F}. The plasma is weakly ionized at the bottom of the chromosphere, but the plasma is almost fully ionized at the top. Therefore, the dominate physical mechanisms in the magnetic reconnection process could vary with height from the bottom to the top of the chromosphere. The numerical results by \citet{2012ApJ...760..109L, 2013PhPl...20f1202L} demonstrate that the recombination is the main effect to cause fast magnetic reconnection in the up chromosphere. Our latest paper \citep{2015ApJ...799...79N} verifies that the plasmoid instability is very important to lead to fast magnetic reconnection at around  the middle of the chromosphere, where the neutral-ion collisional mean free path is smaller than the critical width of the current sheet for secondary islands appearing. The speeds of reconnection out flows are in the range of observed speeds of jets in solar chromosphere.    

Most of the previous numerical simulations which focus on the formations of  EBs  \citep[e.g.,][]{2001ChJAA...1..176C, 2007ApJ...657L..53I, 2009A&A...508.1469A}, chromosphere jets \citep[e.g.,][]{2010A&A...510A.111D, 2011A&A...535A..95D, 2013ApJ...777...16Y} and micro-flares \citep[e.g.,][]{2012ApJ...751..152J, 2014ApJ...788L...2A} are studied based on the single fluid MHD equations with an assumed anomalous resistivity. The topology structures and many characteristics in these numerical simulations can match well with the observation results. For example, the numerical results in the paper by \citet{2001ChJAA...1..176C}, \citet{2009A&A...508.1469A} and \citet{2011RAA....11..225X} show that the temperature enhancement for EBs is around  several thousands Kelvin and the lifetime is around several minutes, which are similar as the observation results of EBs. By setting suitable initial magnetic fields and plasma density, the numerical simulations can lead to the formation of chromosphere jets which have similar topology structures as from the observations, the up-flow speeds from the simulations can be in the range around $10-130$~km$^{-1}$ \citep{2010A&A...510A.111D, 2011A&A...535A..95D, 2013ApJ...777...16Y}, which are the same as the speeds of the observed chromosphere jets. On the other hand, some numerical simulations focus more on the physical mechanisms of magnetic reconnection in solar chromosphere, the partially ionized effects including ambipolar diffusion and recombination have been studied detailedly \citep[e.g.,][]{2012ApJ...760..109L, 2015ApJ...805..134M, 2015PASJ...67...96S}. However, with an initial temperature as low as several thousand Kelvin,  the temperature of the plasma in these chromosphere magnetic reconnection simulations has never been increased by more than one order of magnitude because of high plamsa $\beta$ \citep{2012ApJ...751...56M} or low numerical resolutions. 

In our latest paper\citep{2015ApJ...799...79N}, magnetic reconnection in the middle of the chromosphere ($1000$~km above the solar surface) has been studied carefully. The numerical results in the paper indicate that the initial low temperature plasma ($7000$~K) can be heated to above $8\times10^4$~K by multiple internal slow-mode shocks in the plasmoids inside the current sheet region. In this work, we have built a more realistic model to study the magnetic reconnection in the lower chromosphere (around $100\sim600$km above the solar surface). The more realistic density stratification, radiation cooling, heating, anisotropic heat conduction and ambipolar diffusion effects are all included.  By analyzing the numerical simulation results, we attempt to find out the possible physical mechanism for heating plasma to a high temperature in the lower solar chromosphere/photosphere. The initial conditions and numerical model are described in Section II. We present our numerical results in section III.  Section IV provides a short conclusion and discussion.

\section{Numerical model and initial conditions}
\label{s:model}
As demonstrated in our latest paper, the single-fluid MHD equations  can be used if the collisional coupling between ions and neutrals is strong, which is valid on length-scales exceeding the neutral-ion collisional mean free path 
$\lambda_\mathrm{ni}$. In the lower chromosphere, the coupling of ions and neutrals is even stronger than that in the middle of the chromosphere. According to the initial physical parameters in all the cases given below, we can calculate and find that the current sheet width for onset of the plasmoid instability is several orders of magnitude above $\lambda_\mathrm{ni}$. Therefore, the onset and evolution of plasmoid instability in this part of the solar atmosphere can generally be described using the one-fluid equations as below:

\begin{eqnarray}
 \partial_t \rho &=& -\nabla \cdot (\rho \mathbf{v})                                                                \\ 
 \partial_t \mathbf{B} &=& \nabla \times (\mathbf{v} \times \mathbf{B}-\eta\nabla \times \mathbf{B}+\mathbf{E}_\mathrm{AD})\label{e:induction}\\
  \partial_t (\rho \mathbf{v}) &=& -\nabla \cdot \left[\rho \mathbf{v}\mathbf{v}
                              +\left(p+\frac {1}{2\mu_0} \vert \mathbf{B} \vert^2\right)\mbox{\bfseries\sffamily I} \right]  \nonumber \\
                              & &+\nabla \cdot \left[\frac{1}{\mu_0} \mathbf{B} \mathbf{B} \right] + \rho \mathbf{g} \\
 \partial_t e &=& - \nabla \cdot \left[ \left(e+p+\frac {1}{2\mu_0 }\vert \mathbf{B} \vert^2\right)\mathbf{v} \right] \nonumber\\
       & &+\nabla \cdot \left[\frac {1}{\mu_0} (\mathbf{v} \cdot \mathbf{B})\mathbf{B}\right] + \nabla \cdot \left[ \frac{\eta}{\mu_0} \textbf {B} \times (\nabla \times \mathbf{B}) \right] \nonumber\\
                                       & &-\nabla \cdot \left[\frac{1}{\mu_0}\mathbf{B} \times \mathbf{E}_\mathrm{AD}\right] \nonumber-\nabla \cdot \mathbf{F}_\mathrm{C}\\
                                       & &+\rho \mathbf{g} \cdot \mathbf{v}+\mathcal{L}_\mathrm{rad}+\mathcal{H}   \\
   e &=& \frac{p}{\Gamma_0-1}+\frac{1}{2}\rho \vert  \mathbf{v} \vert^2+\frac{1}{2\mu_0}\vert \mathbf{B} \vert^2          \\
   p &=& \frac{(1+Y_\mathrm{i}) \rho}{m_\mathrm{i}} k_\mathrm{B}T .
\end{eqnarray}

The same as in our previous paper \citep{2015ApJ...799...79N}, $\rho$ is the plasma mass density, $\mathbf{v}$ is the centre of mass velocity, $e$ is the total energy density, $\mathbf{B}$ is the magnetic field, $\eta$ is the magnetic diffusivity, and $p$ is plasma thermal pressure, $\mathbf{g}=-273.9~\mbox{m\,s}^{-2}~\mathbf{e}_y$ is the gravitational acceleration of the Sun. The ambipolar electric field is given by $\mathbf{E}_\mathrm{AD} = \mu_0^{-1}\eta_\mathrm{AD}[(\nabla \times \mathbf{B}) \times \mathbf{B}]\times \mathbf{B}$, where $\eta_\mathrm{AD}$ is the ambipolar diffusion coefficient and we use the same formula as given in equation(22) in our previous paper \citep{2015ApJ...799...79N},
 \begin{eqnarray}
    \eta_\mathrm{AD}&=& 1.65 \times 10^{-11}
  (\frac{1}{Y_i}-1)\frac{1}{\rho^2\sqrt{T}}
                        ~~\mbox{m$^3$\,skg$^{-1}$}.
 \end{eqnarray}
 $\mathcal{L}_\mathrm{rad}$ is the radiative cooling function and $\mathcal{H}$ is the heating function. $Y_\mathrm{i} $ is the ionization degree of the plasma. Comparing with our previous paper \citep{2015ApJ...799...79N}, the anisotropic heat conduction term $\mathbf{F}_\mathrm{C}$ is included in this work. 
 
We still follow the radiative loss in \citet{1990ApJ...358..328G} and set
\begin{eqnarray}
 \mathcal{L}_\mathrm{rad} &=& -1.547\times 10^{-42}\, Y_\mathrm{i} \left(\frac{\rho}{m_\mathrm{i}}\right)^2 \alpha T^{1.5},
\end{eqnarray}
Different from the previous paper, the heating function is chosen linearly depending on the mass density as 
 \begin{eqnarray}
 \mathcal{H}&=& 1.547\times10^{-42}\, Y_\mathrm{i} \frac{\rho \rho_0}{{m_\mathrm{i}}^2} \alpha T_0^{1.5}, 
\end{eqnarray}
where $\rho_0$ and $T_0$ are the initial plasma mass density and temperature separately.

The anisotropic heat conduction term $\mathbf{F}_{C}$ is given by
 \begin{eqnarray}
  \mathbf{F}_{C}=-\kappa_{\parallel}(\nabla T\cdot \hat{\mathbf{B}} )\hat{\mathbf{B}}-\kappa_{\perp}(\nabla T-(\nabla T\cdot \hat{\mathbf{B}})\hat{\mathbf{B}})
\end{eqnarray}
where $\hat{\mathbf{B}}=\mathbf{B}/ \vert \mathbf{B} \vert $ is the unit vector in the direction of magnetic field.  According to the paper by \citet{1961ApJ...134...63O}, the conductivity coefficient $\kappa$ is contributed partially by neutral particles and partially by the charged particles. However, the conductivity coefficient contributed by charged particles is  strongly suppressed in the direction perpendicular to the magnetic field. In this work, we build a conductivity coefficient model according the results presented in Figure.1 in the paper by \citet{1961ApJ...134...63O}. The parallel and perpendicular conductivity coefficient are assumed as below 

\begin{eqnarray}
  \kappa_{\parallel}=1.96\times10^{-11}T^{32/13}+10^{8}T^{-2.5},\\
  \kappa_{\perp} = 10^{8}T^{-2.5}.
 \end{eqnarray}
 
The term $1.96\times10^{-11}T^{\frac{32}{13}}$ is contributed by the charged particles and $10^{8}T^{-2.5}$ is contributed by neutral particles. During the range for $5000$~K $<T< 10^5$~K, the charged particles dominate and the first term is much larger than the second term at the right hand side of equation (11). In the direction perpendicular to the magnetic field, the term contributed by charged particles is totally modified and it is much smaller than the one contributed by neutral particles.  Therefore, we only include the term $10^{8}T^{-2.5}$  in equation (12).
 
 Many papers \citep[e.g.,][]{2009ApJ...697..693J, 2012SoPh..280...51J, 2013A&A...558A..30Q} show that the transverse magnetic fields in the photosphere could also be very strong. 
The observational and numerical results indicate that  the vertical and horizontal current sheet structures can both possibly be formed in the low solar atmosphere  \citep{2004ApJ...614.1099P, 2009ApJ...701.1911P, 2012EAS....55..115P, 2012SoPh..278...73V}. In this work, we present our simulation results in five cases. The horizontal current sheets without gravity and density stratification have been studied in Case~I, Case~II and Case~IIa. One should note that the solar surface is spherical, though the current sheet is horizontal to the solar surface, the bi-directional flows with blue-red shift as suggested by observations can still possibly be observed as long as the current sheet is far from the center of the solar surface. According to the VAL-C chromosphere model \citep{1981ApJS...45..635V}, Case~I can represent a magnetic reconnection process at around 600~km above the solar surface, and the current sheets in Case~II and Case~IIa are at even lower places (about 250~km above the solar surface). The vertical current sheets with density stratification have been studied in Case~III and Case~IIIa, which are located at around $100\sim 600$~km above the solar surface. The important initial parameters and conditions in all the five cases are listed below in table.~1 and in the following subsections. Adaptive mesh refinement (AMR) is applied in our simulations. The same as in the previous paper \citep{2015ApJ...799...79N},  the highest refinement level is 10, which corresponds to a grid resolution $\Delta x \approx 6.1$~m. Convergence studies have been carried out by repeating the simulations with both a lower and a higher resolution to make sure that the numerical diffusion is smaller than the physical magnetic diffusion.

\subsection{Case~I, II and IIa}\label{ss:CasesI+II+IIa}
 We ignore the density stratification effect in Case~I, II and IIa, because the width of the horizontal current sheet in our simulations is much shorter than the length. The simulation domain extends from $x=0$ to $x=L_0$ in $x$ direction and from $y=-0.5L_0$ to $y=0.5L_0$ in $y$ direction in the three cases, with $L_0=10^6$~m. Outflow boundary conditions are used in $x$ direction and Inflow boundary conditions in $y$ direction. For the inflow boundary conditions, the fluid is allowed to flow into the domain but not to flow out; the gradient of the plasma density vanishes; the total energy set such that gradient in thermal energy density vanishes; vanishing gradient of parallel components plus divergence-free extrapolation of the magnetic field. For the outflow boundary conditions, the fluid is allowed to flow out of the domain but not to flow in and the other variables are set by using the same method as the inflow boundary conditions. The horizontal force-free Harris current sheet is used as the initial equilibrium configuration of magnetic fields in Case~I,
\begin{eqnarray}
  B_{x0}&=&-b_0\tanh[y/(0.05L_0)]                               \\
  B_{y0}&=&0   \\
  B_{z0}&=&b_0/\cosh[y/(0.05L_0)].
\end{eqnarray}
The magnetic fields in the low solar atmosphere could be very strong \citep{2014Sci...346C.315P, 2014PhPl...21i2901K, 2015ApJ...812...11V, 2009ApJ...697..693J, 2012SoPh..280...51J} and the magnetic field can exceed $0.1$~T in both the intranetwork and the network quiet region \citep[e.g.,][]{2007ApJ...670L..61O, 2008A&A...477..953M, 2009ApJ...697..693J, 2012SoPh..280...51J}. In the work by \cite{2012SoPh..280...51J}, the maximum of the field strength was found to be $0.15$~T.  The magnetic field could be even stronger in the active region near the sunspot. Therefore, we set $b_0=0.05$~T in Case~I and Case~II,  and $b_0=0.15$~T in Case~IIa. Due to the force-freeness and neglect of gravity, the initial equilibrium thermal pressure is uniform. The initial temperature and plasma density are set as $T_0=4200$~K and $\rho_0=1.66057\times10^{-6}$~kg\,m$^{-3}$ in Case~I, and $T_0=4800$~K and $\rho_0=3.32114\times10^{-5}$~kg\,m$^{-3}$ in Case~II and Case~IIa. Therefore, the initial plasma $\beta$ is calculated as $\beta\simeq0.0583$ in Case~I,  $\beta \simeq 1.332$ in Case~II and $\beta \simeq 0.148$ in Case~IIa. The initial ionization degree is assumed as $Y_\mathrm{i}=10^{-3}$ in Case~I,  and $Y_\mathrm{i}=1.2\times10^{-4}$ in Case~II and IIa. The magnetic diffusion in this work matches the form computed from the solar atmosphere model in \citet{2012ApJ...747...87K},  and we set $\eta=\left[5\times 10^4 (4200/T)^{1.5}+1.76\times10^{-3}T^{0.5}Y_i^{-1}\right]$~m$^2$\,s$^{-1}$ in Case~I, and $\eta=\left[5\times 10^4 (4800/T)^{1.5}+1.76\times10^{-3}T^{0.5}Y_i^{-1}\right]$~m$^2$\,s$^{-1}$ in Case~II and IIa. The first part $\sim T^{-1.5}$ is contributed by collisions between ions and electrons, the second part $\sim T^{0.5} Y_i^{-1}$ is contributed by collisions between electrons and neutral particles. Small perturbations for both magnetic fields and velocities at $t=0$ make the current sheet to evolve and secondary instabilities start to appear later in the three cases. The forms of perturbations are listed as below:
 \begin{eqnarray}
   b_{x1}&=&-pert\cdot b_0 \cdot \sin\left(2\pi \frac{y+0.5L_0}{L_0}\right) \cdot \cos\left(2\pi \frac{x+0.5L_0}{L_0}\right)    \\
   b_{y1}&=&pert\cdot b_0 \cdot \cos\left(2\pi \frac{y+0.5L_0}{L_0}\right) \cdot \sin\left(2\pi \frac{x+0.5L_0}{L_0}\right) \\
   v_{y1}&=&-pert\cdot v_{A0} \cdot \sin\left(\pi \frac{y}{L_0}\right) \cdot \frac{random_n} {Max\left(\arrowvert random_n\arrowvert\right)},
  \end{eqnarray}
 where $pert=0.08$, $v_{A0}$ is the initial Alfv\'en velocity, $random_n$ is the random noise function in our code,  and $Max(\arrowvert random_n\arrowvert)$ is the maximum of the absolute value of the random noise function. This random noise function makes the initial perturbations for velocity in y-direction to be asymmetric, and such an asymmetry makes the current sheet gradually to become more tilted, especially after secondary islands appear. Reconnection process is not really symmetrical in nature \citep{2012ApJ...751...56M}, this is one of the reasons that we use such an noise function. Another reason is that the asymmetric noise function makes the secondary instabilities to develop faster. Figure~1(a) shows the distributions of the current density and magnetic fields at $t=0$ in case~I.   
 
\subsection{Case~III and IIIa}\label{ss:CasesIII+IIIa}
 For the situation with density stratification, we have simulated two cases, Case~III and IIIa. The simulation domain extends from $x=-0.5L_0$ to $x=0.5L_0$ in $x$ direction and from $y=0$ to $y=0.5L_0$ in $y$ direction in these two cases, also with $L_0=10^6$~m. Open boundary conditions are used in $x$ and $y$ directions. The velocity gradient which is perpendicular to the boundary layer is set to zero and the other variables are set by using the same method as the inflow and outflow boundary. The vertical current sheet with initial equilibrium magnetic fields set as,
\begin{eqnarray}
  B_{x0}&=&0                              \\
  B_{y0}&=&b_0\tanh[x/(0.05L_0)]    \\
  B_{z0}&=&b_0/\cosh[x/(0.05L_0)],
\end{eqnarray} 
 with $b_0=0.05$~T in Case~III and $b_0=0.15$~T in Case~IIIa. The initial gas pressure is then give by
   \begin{eqnarray}
    \partial_x p_0(y) &=& 0,  \\
    \partial_y p_0(y) &=& -273.9\rho_0(y), \\
    p_0(y) &=& \frac{(1+Y_\mathrm{i0}) \rho_\mathrm{H0}(y)}{m_\mathrm{i}} k_\mathrm{B}T_0,
 \end{eqnarray}
where, for simplicity, the initial temperature $T_0=5000$~K is chosen to be uniform in both of the two cases, and $\rho_\mathrm{H0}=n_\mathrm{H0}m_i$. According to Equation~(12) in \citet{1990ApJ...358..328G}, the ionization degree is simplified as
 \begin{eqnarray}
   Y_\mathrm{i0}=\frac{n_\mathrm{e0}}{n_\mathrm{H0}} \approx \sqrt{\frac{\phi_0}{n_\mathrm{H0}}},
 \end{eqnarray}
where $\phi_0 \simeq 1.555 \times 10^{15}$~m$^{-3}$ and $\phi_0 \ll n_\mathrm{H0}$ in our model. From Equations~(19--24), we obtain
 \begin{eqnarray}
  \rho_\mathrm{H0} \simeq \rho_\mathrm{H00}\exp\left(-\frac{6.589(y+L_0)}{L_0}\right), 
 \end{eqnarray}
where we assume $\rho_\mathrm{H00}=0.04151425$~kg\,m$^{-3}$. Then the initial ionization degree is calculated as $Y_i\simeq2.13\times10^{-4}$ at $y=0$ and $Y_i\simeq1.10\times10^{-3}$ at $y=0.5L_0$. The values of the important parameters in the two cases at $y=0$, $y=0.25L_0$ and $y=0.5L_0$ are also  compiled in Table~1. The only difference between Case III and IIIa is the value of the magnetic fields, which makes the corresponding plasma $\beta$ in Case~IIIa is nine times smaller than that in Case~III.  Unlike Case~I, Case~II and Case~IIa, the initial symmetric perturbations for velocities are applied in the simulations of Case~III and Case~IIIa. The distributions of magnetic fields and plasma density at beginning in Case~III are presented in Figure~1(b).
 
\begin{table*}
 \caption{Important parameters of the current sheet region in Case I, II, IIa, III and IIIa. Initial values of $T_0$ -- temperature, $n_\mathrm{H0}$ -- hydrogen density, $b_0$--initial magnetic field, $ \beta$ -- initial plasma $\beta$, $v_\mathrm{A0}$ -- Alfv\'en velocity,  $S_0$ -- Lundquist number. The height dependence of these parameters is significant in Case III and IIIa.}
\label{Parameter}
  \begin{tabular}{lccccccc}
   \hline
                    &$T_0$(K)& $n_\mathrm{H0}$(m$^{-3}$) & $b_0$(T) & $\beta$&  $v_\mathrm{A0}$(m/s) &$t_{A0}(s)$&$S_0$      \\
    \hline
    Case I     &$4200$ &$              10^{21}$& $0.05$& 0.058&$3.46\times10^4$&$28.9$&$6.92\times10^5$  \\ 
   \hline
    Case II    &$4800$ & $2\times10^{22}$& $0.05$& 1.332&$7.74\times10^3$&$129.2$&$1.55\times10^5$  \\ 
    \hline
    Case IIa  &$4800$ &$2\times 10^{22}$&$0.15$& 0.148&$2.32\times10^4$&$43.1$&$4.64\times10^5$  \\ 
    \hline   
     Case III\\
      ($y=0$)             &$5000$ &$3.44\times 10^{22}$&$0.05$&2.397&$5.90\times10^3$&$169.4$&$1.18\times10^5$ \\                     
      ($y=0.25L_0$)&$5000$ &$6.62\times 10^{21}$&$0.05$&0.460 &$1.34\times10^4$&$74.4$&$2.69\times10^5$ \\
      ($y=0.5L_0$)  &$5000$ &$1.28\times 10^{21}$&$0.05$& 0.089&$3.06\times10^4$&$32.6$&$6.13\times10^5$  \\
     \hline   
     Case IIIa\\
      ($y=0$)             &$5000$ &$3.44\times 10^{22}$&$0.15$&0.265 &$1.77\times10^4$&$56.5$&$3.54\times10^5$ \\                     
      ($y=0.25L_0$)&$5000$ &$6.62\times 10^{21}$&$0.15$&0.051 &$4.03\times10^4$&$24.8$&$8.07\times10^5$ \\
      ($y=0.5L_0$)  &$5000$ &$1.28\times 10^{21}$&$0.15$& 0.010&$9.19\times10^4$ &$10.9$&$1.84\times10^6$ \\
    \hline
  \end{tabular}   
\end{table*}

\section{Numerical Results}\label{s:results}
\subsection{Plasma heating in horizontal current sheets }\label{ss:heating}
  Figure~2 shows that the bi-directional reconnection out-flows always exist in the whole magnetic reconnection process, the highest velocity in Case~I can reach around $50$~km\,s$^{-1}$. Figure~3(a) and Figure~7 show that the temperature slowly increases before secondary islands appear. As secondary islands start to appear, it increases sharply and the maximum temperature exceeds $10^5$~K in Case~I. The first panel of Figure~4 shows the distributions of current density and magnetic fields at $t=1.592t_{A0I}$ in Case~I. One can see that different sizes of plasmoids appear, the bigger islands are usually formed by coalescence of smaller islands. In order to see more details inside the plasmoids, we zoom into a small scale as shown in the second panel of Figure~4. The black arrows in this panel represent the velocities of plasmas and the red-blue color contours represent the current density. One can find that the two magnetic islands with comparable size in the merging process are located in the left part of this panel. Many turbulent structures appear at the head of plasmoids and in the coalescence process. Another island is located in the right part of this panel and a relatively much smaller island is located around the middle of this panel. All these islands are moving toward x-direction. As shown in the second and the third panels of Figure~4,  a pair of  slow-mode shocks are formed behind the moving magnetic island which is located in the right of this panel. The temperature is strongly increased at the shock regions and reaches a maximum value. The slow mode shocks with hot structures are disrupted when the two magnetic islands merges into one big island as shown in the left part of the third panel in Figure~4. As we continue to zoom in to the region where the small island locates in the middle of the second panel in Figure~4, we can also find that a pair of small slow-mode shocks behind this small island and the temperature is higher there. Many different sizes of  slow-mode shocks are found at the edges of different sizes of plasmoids.  The fourth panel of Figure~4 shows the distributions of the Mach number, one can find that the maximum Mach number is above 2, which indicates that the strong fast mode shocks are formed below these high Mach number regions.  We use the same method as \citet{2015Sci...350.1238C} to calculate the Mach number, $M_{A}=\frac{v}{\sqrt{v_{sound}^2+v_{A}^2}}$, where $v_{sound}$ is the local sound speed, $v_{A}$ is the local Alfv\'en speed. The temperature increases also appear at these fast-mode shock regions as shown in Figure~3(a) and Figure~4.  However,  they are not as obvious as the temperature increases at the slow-mode shock regions.  
 
 Figure~6(a) shows the total generated thermal energy and Joule heating $\mu_0\eta J^2$,  in the domain $0.2L_0<x<0.8L_0$ and $-0.01L_0<y<0.01L_0$,  during the period $1.352t_{A0I}<t<1.631t_{A0I}$ in Case~I. We should point out that the unit for the variables in Figure~6 is J\,m$^{-1}$, because the energy density is not integrated in z-direction. The method to calculate the total generated thermal energy is the same as that in our previous papers \citep{2012ApJ...758...20N, 2015ApJ...799...79N}. The thermal energy flowing into this region through the boundaries ($x=0.2L_0, x=0.8L_0, y=-0.01L_0$ and $y=0.01L_0$) from $t=1.352t_{A0I}$ to time t is denoted as $ETF(t)$, and it is calculated as:
\begin{eqnarray}
 ETF(t) & = & \int_{1.352t_{A0I}}^{t}\!\int_{-0.01L_0}^{0.01L_0} (\Gamma_0\varepsilon(0.2L_0,y,t)v_x(0.2L_0,y,t)+F_{Cx}(0.2L_0,y,t)) dy dt-
                    \nonumber\\
                    &&\int_{1.352t_{A0I}}^{t}\!\int_{-0.01L_0}^{0.01L_0} (\Gamma_0\varepsilon(0.8L_0,y,t)v_x(0.8L_0,y,t)+F_{Cx}(0.8L_0,y,t)) dy dt+
                     \nonumber\\
                      & &\int_{1.352t_{A0I}}^{t}\!\int_{0.2L_0}^{0.8L_0} (\Gamma_0\varepsilon(x, -0.01L_0,t)v_y(x, -0.01L_0,t)+F_{Cy}(x, -0.01L_0,t)) dx dt-
                      \nonumber\\
                     & & \int_{1.352t_{A0I}}^{t}\!\int_{0.2L_0}^{0.8L_0} (\Gamma_0\varepsilon(x, 0.01L_0,t)v_y(x, 0.01L_0,t)+F_{Cy}(x, 0.01L_0,t)) dx dt,
\end{eqnarray}
where $F_{Cx}$ and $F_{Cy}$ are the heat conduction terms as shown in equation(10) in x-direction and y-direction, respectively. Note that this quantity may have negative signs if thermal energy flows out of the region. The thermal energy confined to this region at time t is denoted as $ETL(t)=\int_{-0.01L_0}^{0.01L_0}\!\int_{0.2L_0}^{0.8L_0} \varepsilon(x,y,t) dx dy$. The initial thermal energy at $t=1.352t_{A0I}$ is denoted as $ETI$. The total radiated thermal energy from $t=1.352t_{A0I}$ to time t is denoted as $ERAD(t)$. Therefore, in this domain, the generated thermal energy at time $t$ is $ETG(t) = ERAD(t)+ETL(t)-ETF(t)-ETI$. In Figure~6(a), one can find that the total generated thermal energy is much higher than the thermal energy by Joule heating. Only $2.38\%$ of the generated thermal energy is contributed by Joule heating during this period and this domain. The original raw data calculated from NIRVANA code can only be transformed to uniform IDL data and the highest level (level 10) IDL data in the domain $0.2L_0<x<0.8L_0$ and $-0.01L_0<y<0.01L_0$ are too big to be analyzed. Therefore, we only analyze the level 6 IDL data to get the plots in Figure~6(a) and the energy by Joule heating could be underestimated. However, we have zoomed in to several much smaller regions to calculate the Joule heating during a fixed period by using the highest resolution, and we still find that the Joule heating is much lower than the total generated thermal energy as long as the shocks are included in these small regions. Figure~6(b) shows the total generated thermal energy and the Joule heating calculated by using level 10 IDL data,  in the domain $0.55L_0<x<0.60L_0$ and $-0.002L_0<y<0.006L_0$ during the period $1.330t_{A0I}<t<1.609t_{A0I}$. One can still find that only $6.98\%$ of the total generated thermal energy is contributed by Joule heating during this period. Therefore, we can conclude that Joule heating is not the main physical mechanism to generate thermal energy in Case~I. The slow-mode and fast-mode shocks dominate for heating plasma to high temperatures. 
  
 The initial plasma density in Case~II is 20 times higher than that in Case~I. As shown in Figure~3(b) and Figure~7(a), the highest temperature can only increase from $4800$~K to around $8400$~K in Case~II. The temperature exceeds $6400$~K in most areas of the reconnection regions at $t=2.093t_{A0II}$. The highest temperature also does not appear at the magnetic reconnection X-points, but inside the plasmoids. As we zoom in to the small regions, the slow-mode and fast-mode shocks can also be identified at the edges of some plasmoids in Case~II, as shown in Figure~5. However, the temperature increases at these shock regions are much less obvious than those in Case~I. The dynamic structures are relatively much smoother than those in Case~I, and the plasmas are also relatively less turbulent in the coalescence process. Since the maximum Mach number is only around 1, the fast-mode shocks are much weaker than those in Case~I. Figure~6(c) presents the total generated thermal energy and the Joule heating $\mu_0\eta J^2$,  in the domain $0.2L_0<x<0.8L_0$ and $-0.01L_0<y<0.01L_0$,  during the period $1.850t_{A0II}<t<2.186t_{A0II}$ in Case~II. Though the level 6 IDL data could possibly make the Joule heating underestimated in Figure~6(c), one can still find that $\sim87\%$ of the total generated thermal energy is contributed by Joule heating during this period. Therefore, we can conclude that Joule heating is the main physical mechanism to generate thermal energy in Case.~II. 
  
The initial plasma density and temperature in Case~IIa is the same as those in Case~II, but the initial magnetic fields set in Case~IIa is three times higher than that in Case~II. As shown in Figure~3(c) and Figure~7, the maximum temperature can exceed $4\times10^4$~K in Case~II after the secondary islands appear.  The slow-mode and fast-mode shocks also dominate for generating thermal energy in Case~IIa. 
 
As described in Section~II,  Case~I can represent a magnetic reconnection process at around 600~km above the solar surface, and the current sheets in Case~II and Case~IIa are at even lower places (about 250~km above the solar surface). The magnetic fields must be strong enough for heating plasma to high temperatures in such a high plasma density and low temperature environment.  For the initial plasma density as high as $2\times10^{22}$~m$^{-3}$ in the photosphere, the temperature can be heated above $4\times10^4$~K in a magnetic reconnection process with initial magnetic fields $b_0=0.15$~T (as shown in Case~IIa). We have run an additional case which is not shown in this work, the initial magnetic field is set with $b_0=0.2$~T, and the other parameters are the same as those in Case~IIa. The maximum temperature reaches around $8\times10^4$~K in this extreme case. However, we should point out that such a strong magnetic field with $b_0=0.2$~T is rarely observed in the photosphere. The slow-mode and fast-mode shocks attached at the edges of the multiple level plasmoids in a magnetic reconnection process with low plasma $\beta$ are the main physical mechanisms for generating the high temperature plasmas. If the magnetic fields are weak and the plasma $\beta$ is high,  the temperature in a magnetic reconnection process can only be increased for several thousand Kelvin, Joule heating is the main mechanism for heating plasma. 

\subsection{Gravity effect in vertical current sheets}\label{ss:gravity}

Figure~3(d)  shows the distributions of temperature in the current sheet region in Case~III at $t=1.832t_{A0IIIU}$, $t=2.599t_{A0IIIU}$ and $t=3.182t_{A0IIIU}$, where $t_{A0IIIU}=32.6$~s is the initial Alfv\'en time at the top boundary $y=0.5L_0$. The light blue arrows in this figure represent the velocity distributions of plasma. As shown in the first panel of Figure~3(d), we can see both the up flows and down flows at $t=1.832t_{A0IIIU}$ in the reconnection region before secondary islands appear. However, the gravity effect is so strong in such a low atmosphere that all the secondary islands eventually move downward in the negative y-direction. The temperature strongly increases in front of these downward plasmoids, the maximum  temperature can reach 16000~K before the plasmoids move out the simulation domain. The Mach number as shown in Figure~8(a) is larger than 1 at some regions inside the current sheet.  Figure~8(b) shows that the positive peaks of $B_x$ and the peaks of temperature along y direction at $x=0$ are almost at the same positions. Comparing Figure~8(a) and Figure~8(b), one can find that these peaks are all located below the regions with relatively large Mach number. These phenomena indicate that the small scale fast mode shock-fronts are at these peaks. Since the strong gravity effect prevents the current sheet becoming thinner, we can only see the secondary plasmoids and no higher order plasmoids appear. The slow-mode shocks are also not  formed in Case~III. Figure~6(d) shows the total generated thermal energy and the Joule heating in the domain $-0.01L_0<x<0.01L_0$ and $0<y<0.1L_0$ during the period $2.850t_{A0IIIU}<t<3.300t_{A0IIIU}$.  Joule heating contributes around $27\%$ of the total generated thermal energy during this period inside this domain, the rest is contributed by those fast-mode shocks. 

Though the initial reconnection up flows in Case~IIIa is around three times higher than that in Case~III, the gravity effect in the negative y-direction still makes the up flow velocities decrease with time.  Only some small secondary plasmoids can flow out the up boundary of the simulation domain. The coalescent  bigger islands are too heavy to flow out and they drop downward in negative y-direction eventually. As shown in Figure~7(b), the maximum temperature can reach around $2.8 \times10^4$~K in the reconnection region in Case~IIIa.  

In the observed reconnection events which relate with EBs, IRIS bombs or jets, the current sheet regions are usually tilted in the atmosphere, they are not exactly horizontal or vertical to the solar atmosphere. One can expect that the plasma can possibly be heated to much higher temperatures observed by IRIS in a more realistic tilted current sheet that includes the horizontal component.

\section{Conclusions and discussions}\label{s:discussion}
In this work, we have studied the magnetic reconnection process in the low solar chromosphere by using 2.5 dimensional MHD model. The simple radiative cooling and heating terms \citep{1990ApJ...358..328G}, heat conduction \citep{1961ApJ...134...63O}, ambipolar diffusion, effects of partial ionization, and gravity are all included in our model. The magnetic diffusion in our simulations are close to the realistic form given by \cite{2012ApJ...747...87K}. The adaptive mess refinement makes the resolution to be high enough and the numerical diffusion to be smaller than the physical magnetic diffusion.  By analyzing the numerical results in different cases, we can make the main conclusions as below:

(1) The high temperature explosions ($\gtrsim 8\times10^4$~K) observed by IRIS can be formed in a magnetic reconnection process at around the temperature minimum region in the low solar atmosphere, and they can even be formed in the photosphere with extreme strong magnetic fields ($\sim0.2$~T). The plasma $\beta$ controlled by plasma density and magnetic fields is one important factor to decide how much the plasma can be heated up. The plasma can be heated to exceed $10^5$~K in a low plasma $\beta$ horizontal current sheet with $b_0=0.05$~T, $n_{H0}=10^{21}$~m$^{-3}$, $T_0=4200$~K and $\beta\simeq0.058$. One can expect that the maximum temperature will be increased to an higher value for an even lower plasma $\beta$ case.  The slow-mode and fast-mode shocks present at the edges of the well developed plasmoids inside the current sheet regions are the main mechanisms to heat plasma and generate the high temperature explosions.

(2) The low temperature events ($\sim 10^4$~K) are formed in a magnetic reconnection process with high plasma $\beta$ and weak magnetic fields in the low solar atmosphere. Joule heating is the main mechanism to heat plasma and the temperature increases are at most only several thousand Kelvin. The maximum temperature can only reach around $8400$~K in a horizontal current sheet with $b_0=0.05$~T, $n_{H0}=2\times10^{22}$~m$^{-3}$, $T_0=4800$~K and $\beta\simeq1.332$. The plasma $\beta$ in the low temperature events is larger than $1$ and at least one magnitude higher than that in the high temperature hot explosions ($\gtrsim 8\times10^4$~K). 
 
(3) Gravity in the low solar atmosphere can strongly hinder the plasmoind instability and the formation of slow-mode shocks in a vertical current sheet. Only small secondary islands are formed; these islands, however, are not well developed as those in the horizontal current sheets and no higher order plasmoids are formed. The  up-flow speed in the simulation domain decreases with time because of gravity.    

For the first time, our numerical simulations show that the plasma can be heated from $4200$~K to exceed $10^5$~K in a magnetic reconnection process in the low chromosphere at around the temperature minimum region. These results can be applied for understanding the heating mechanisms in the low solar atmosphere through reconnection at current sheets.  Our work could possibly be extended  to explain the formation of common low temperature EBs ($\sim 10^4$~K) and the high temperature IRIS bombs ($\gtrsim 8\times10^4$~K) in the future. However, the ionization non-equilibrium effect is not included in the simulations. Since the hydrogen gas will be totally ionized as the temperature reaches $10^5$~K, the ionization process for the hydrogen gas will consume a lot of thermal energy. This effect will generally decrease the temperature of the hot plasmas. In the paper by \citet{2011RAA....11..225X}, they have studied magnetic reconnection processes in the low solar atmosphere by including the ionization non-equilibrium and the temperature increases in their simulations are only several thousand Kelvin. However, we should point out that the assumed anomalous resistivity is large and the resolutions are low in their work. Therefore, the secondary plasmoid instabilities associated with many small scale shock structures are not resolved in their paper. On the other hand, the initial magnetic fields set in their simulations are around one magnitude lower than those in our simulations. These two factors are probably the main reasons to cause the high temperature plasmas to be not generated in their simulations. We only use the simple Harris Sheet magnetic structures to study the magnetic reconnection process. The observed EB like and IRIS bomb Events have more complicated structures for magnetic fields. For the further studies, we plan to use the two-fluid MHD model for including the ionization non-equilibrium effect, and the more realistic magnetic structures in a reconnection event will be applied in our future work.  

 \acknowledgments
 We would like to thank the team of the ISSI workshop on "Solar UV bursts- a new insight to magnetic reconnection" Lead by Peter Young for fruitful discussions and ISSI for its financial support. Lei Ni would like to thank Hui Tian and Zhi Xu for their helpful discussions. This research is supported by NSFC (Grant No. 11573064),  NSFC (Grant No. 11203069), NSFC grant 11273055, NSFC grant 11333007, the Western Light of Chinese Academy of Sciences 2014, the key Laboratory of Solar Activity grant  KLSA201404, the Program 973 grants 2013CBA01503,  CAS grant XDB09040202, and Special Program for applied Research on Supper Computation of the NSFC-Guangdong Joint Fund (nsfc2015-460, nsfc2015-463, the second phase). We have used the NIRVANA code v3.6 developed by Udo Ziegler at the Leibniz-Institut f\"ur Astrophysik  Potsdam. The authors gratefully acknowledge the computing time granted by the Yunnan Astronomical Observatories and the National Supercomputer Center in Guangzhou , and provided on the facilities at the Supercomputing Platform , as well as the help from all faculties of the Platform.

\clearpage

 \begin{figure*}
\centerline{\includegraphics[width=0.45\textwidth, clip=]{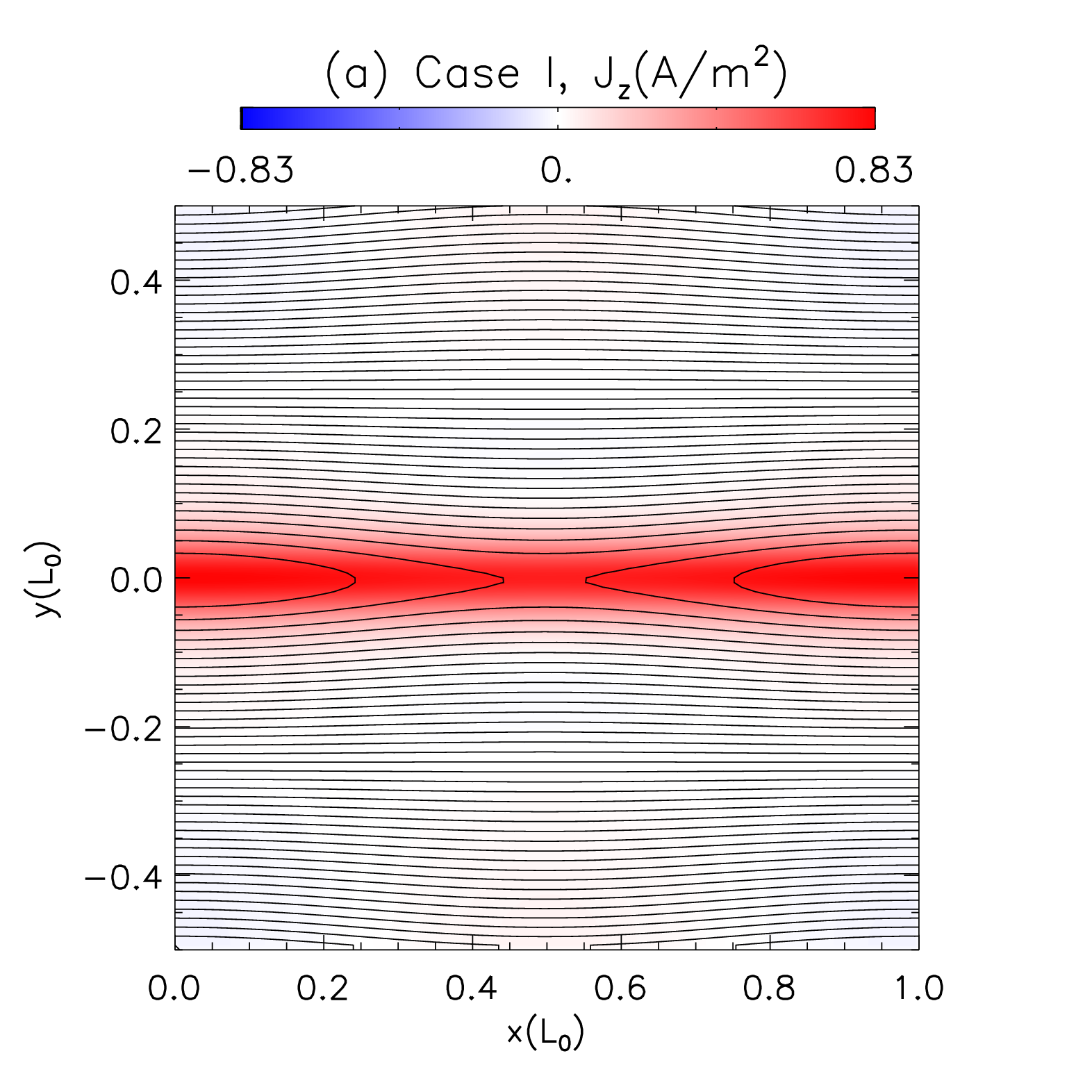}
                    \includegraphics[width=0.45\textwidth, clip=]{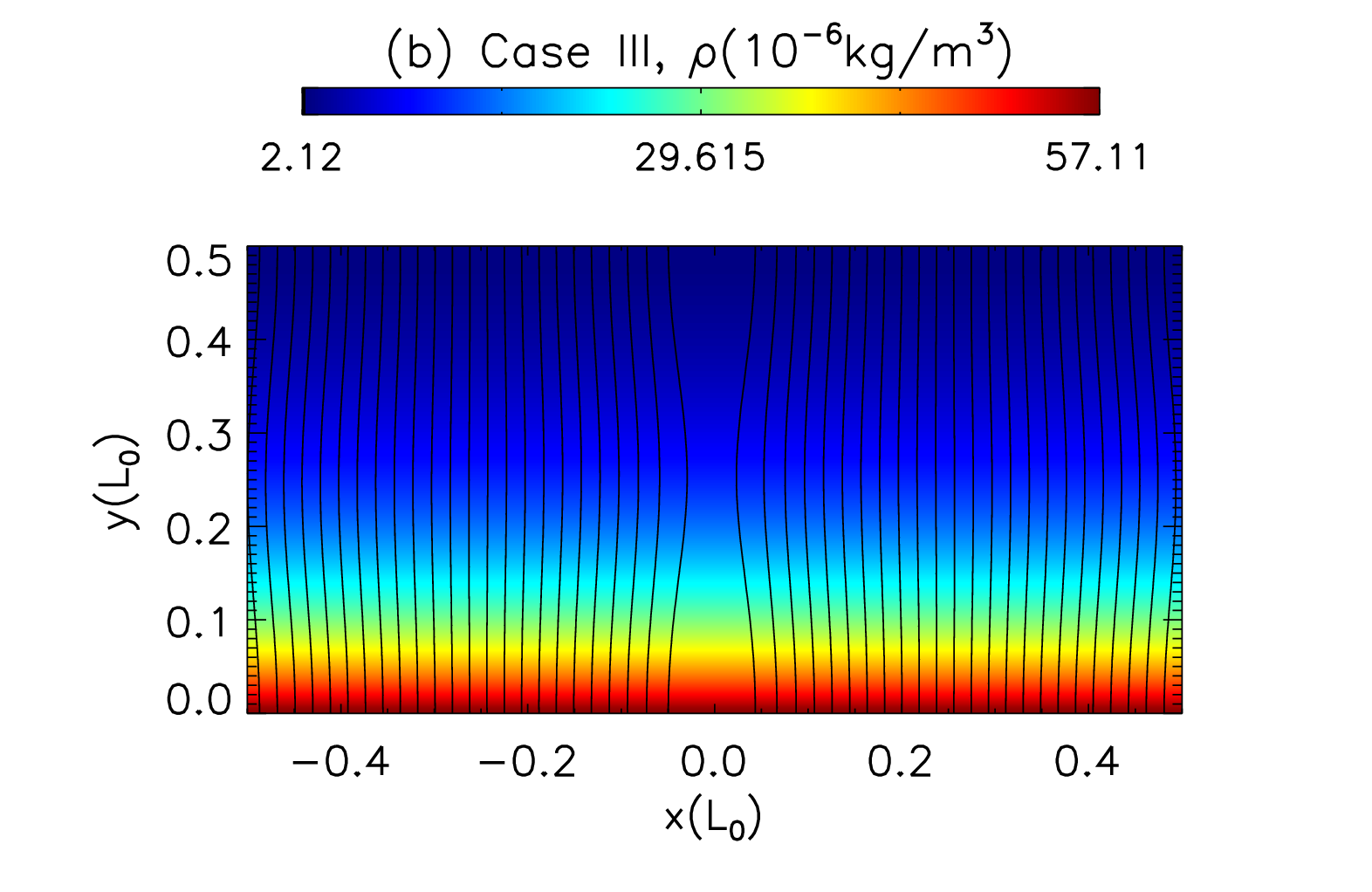}}
    \caption{(a) The distributions of field lines and current density $J_z$ (background color) at $t=0$ in the whole simulation domain for Case.~I.
             (b) The distributions of field lines and plasma density $\rho$ (background color) at $t=0$ in Case.~III.}
\label{fig.1}
\end{figure*}

 \begin{figure*}
\centerline{\includegraphics[width=0.7\textwidth, clip=]{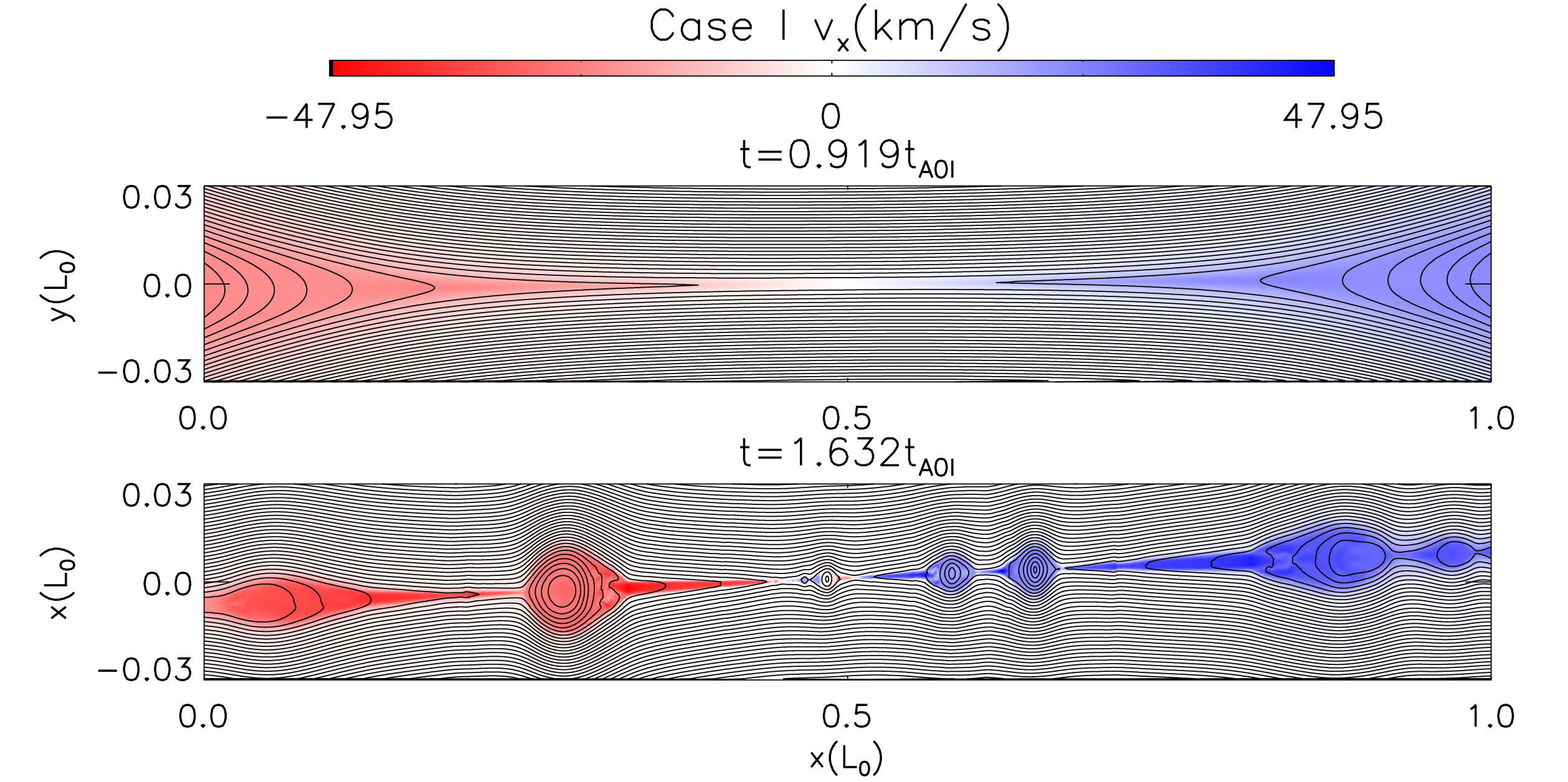}}
    \caption{(a) The distributions of field lines and velocity in x-direction ($v_x$) at $t=0.919t_{A0I}$ and $t=1.632t_{A0I}$ in Case.~I.}
\label{fig.2}
\end{figure*}

\begin{figure*}
  \centerline{\includegraphics[width=0.45\textwidth, clip=]{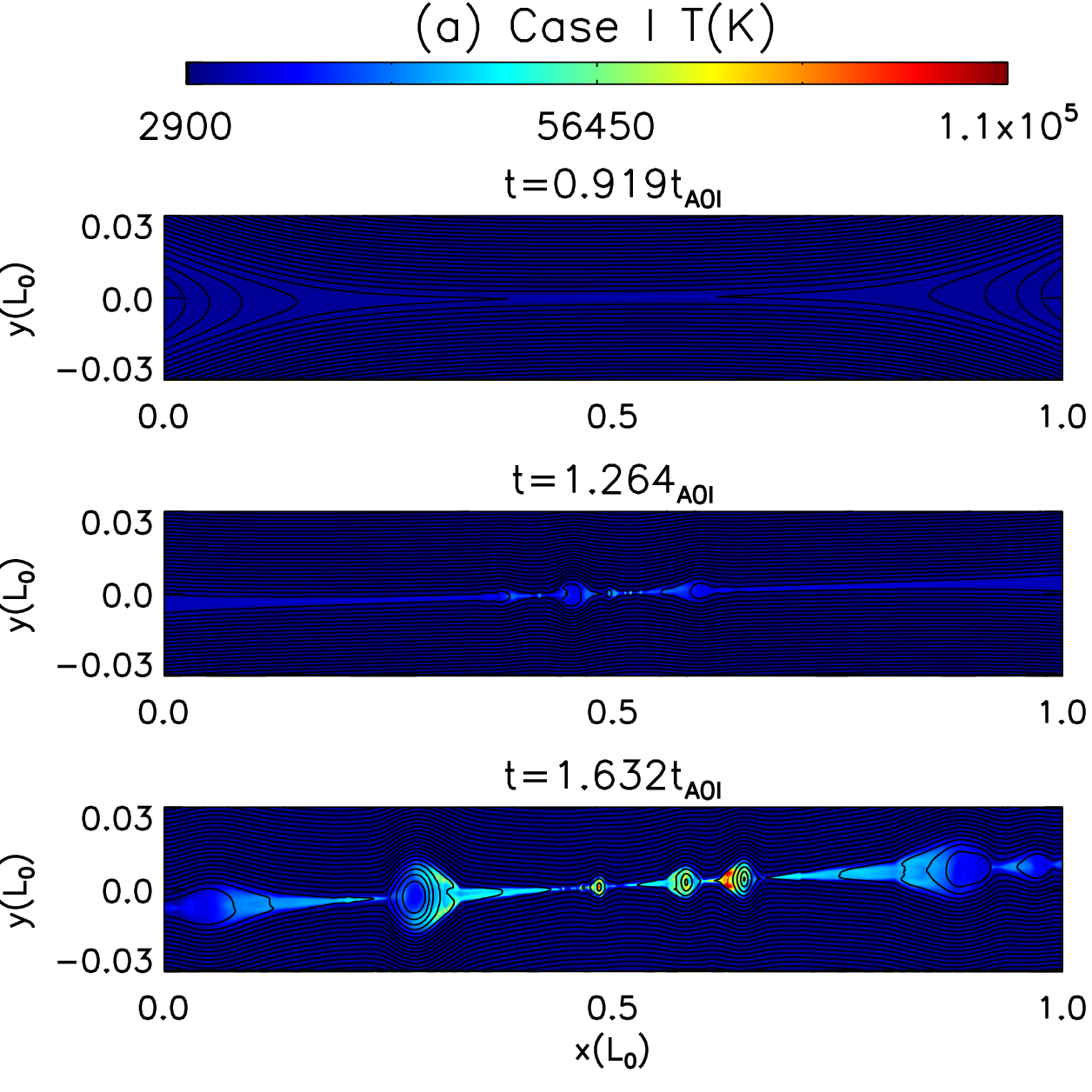}
                        \includegraphics[width=0.45\textwidth, clip=]{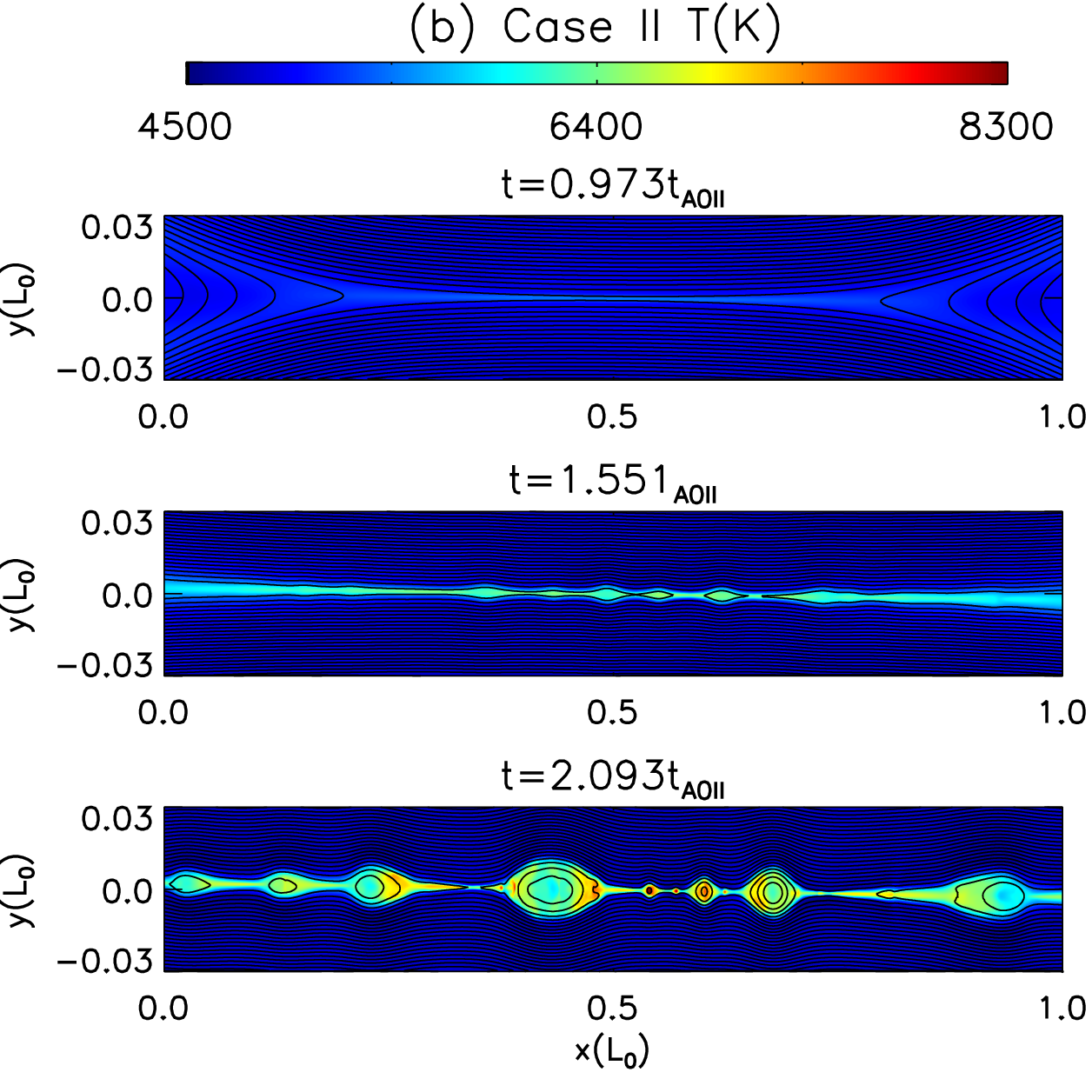}}
    \centerline{\includegraphics[width=0.45\textwidth, clip=]{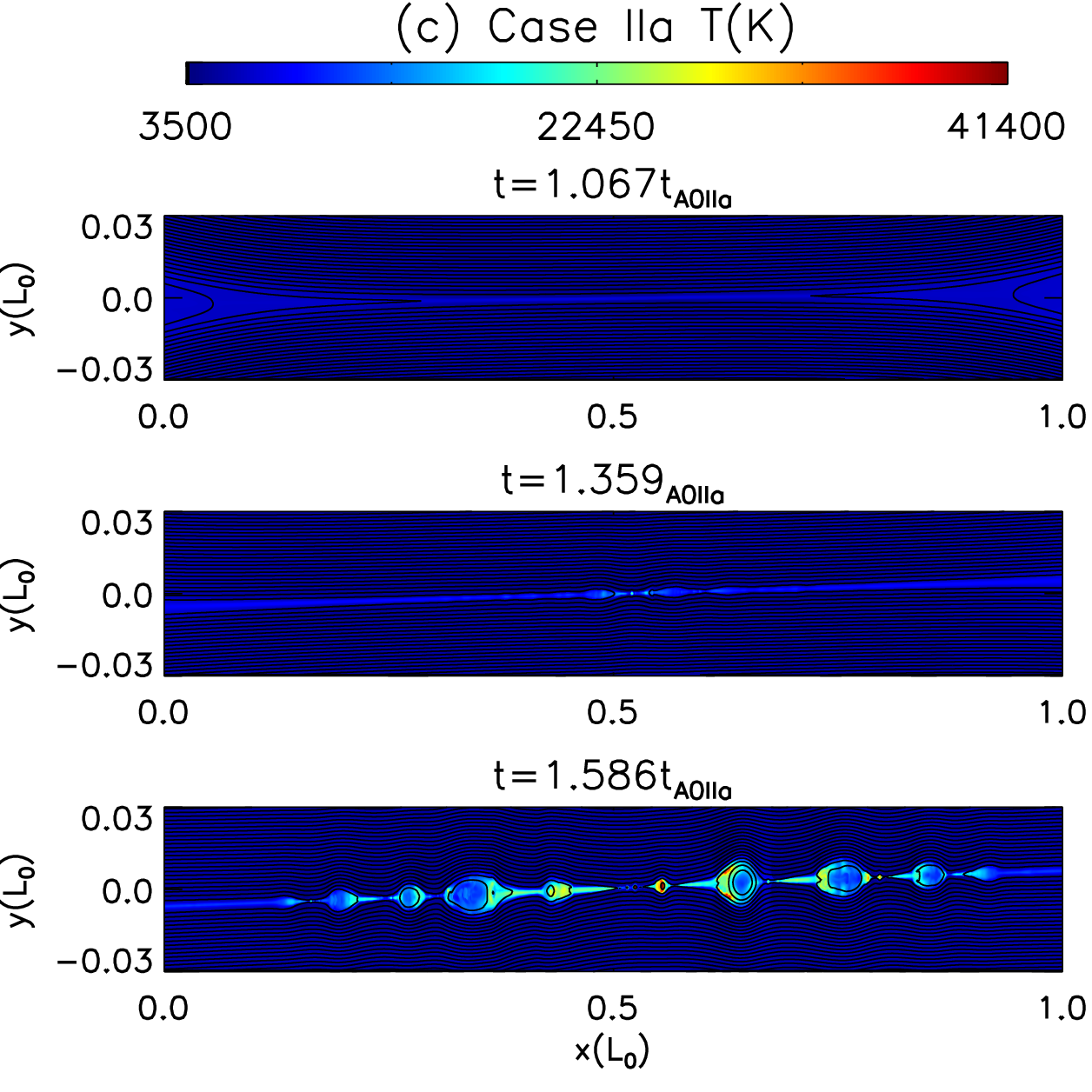}
                        \includegraphics[width=0.45\textwidth, clip=]{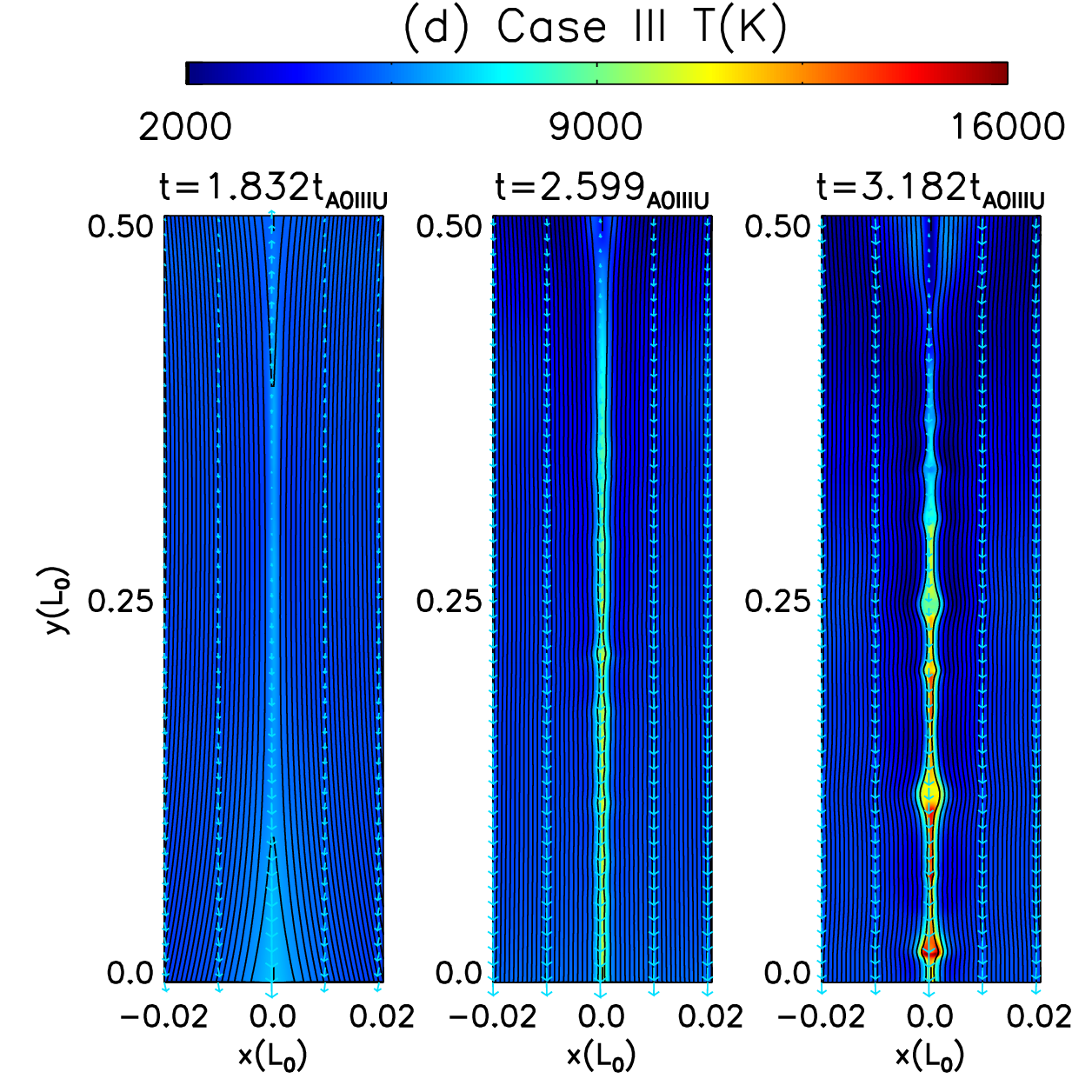}}
   \caption{The distributions of magnetic fields and temperature at three different times,  (a) Case.~I (b) Case.~II, (c) Case.~IIa and (d) Case.~III.  $t_{A0I}$,  $t_{A0II}$, $t_{A0IIa} $ and $t_{A0III}$ are the initial Alfv\'en time in Case.~I,  Case.~II, Case.~IIa and in Case.~III at the up boundary, respectively.}
  \label{fig.32}
\end{figure*}

\begin{figure*}
 \centerline{\includegraphics[width=0.80\textwidth, clip=]{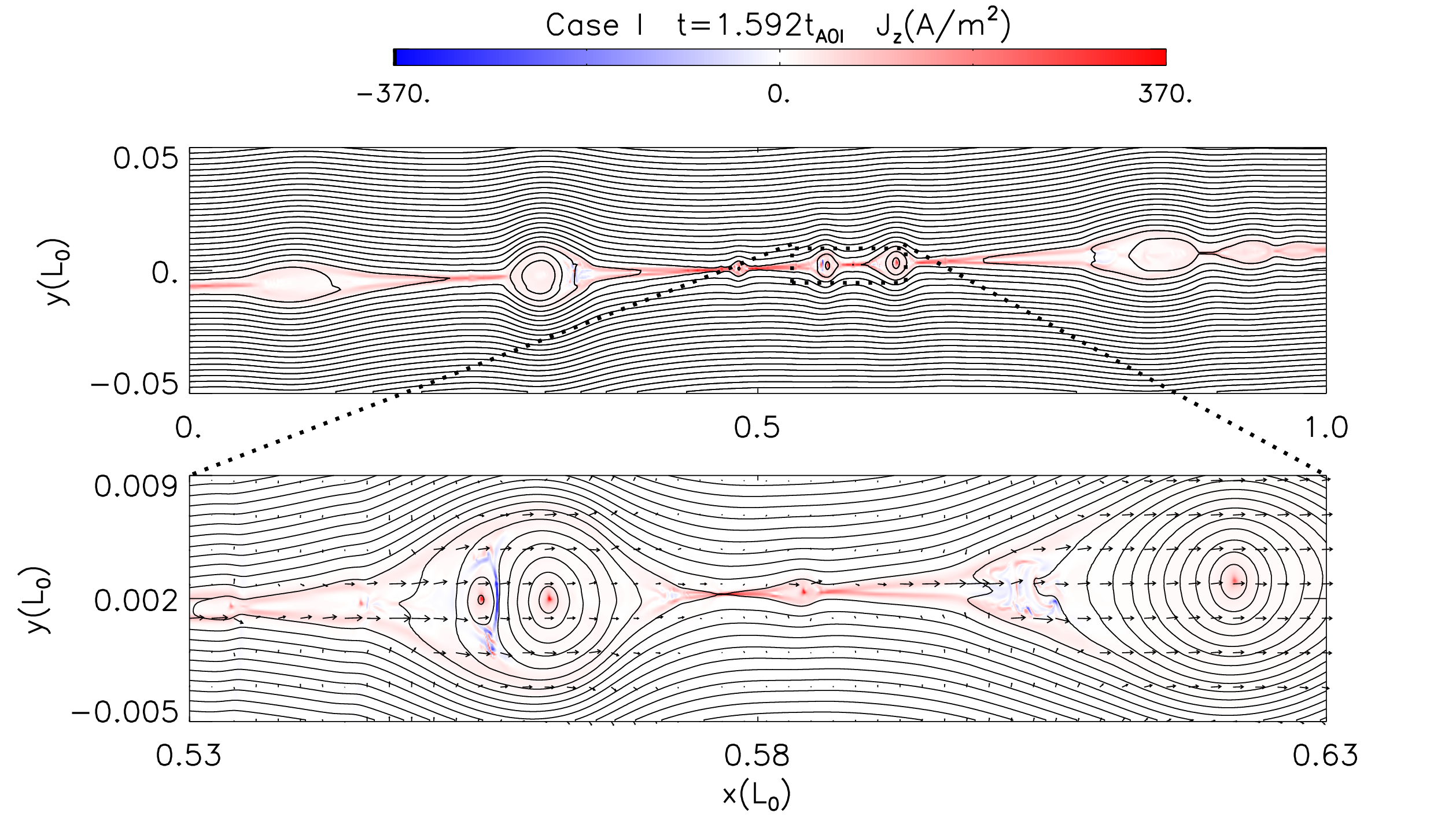}}
  \centerline{\includegraphics[width=0.80\textwidth, clip=]{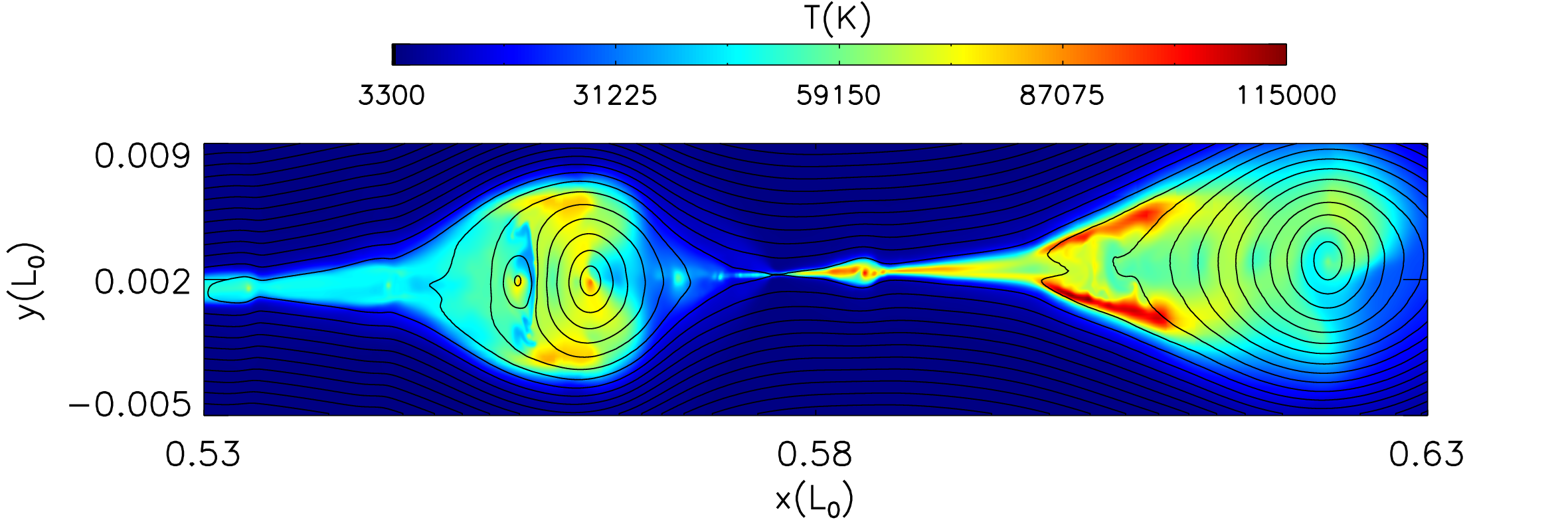}}
 \centerline{\includegraphics[width=0.80\textwidth, clip=]{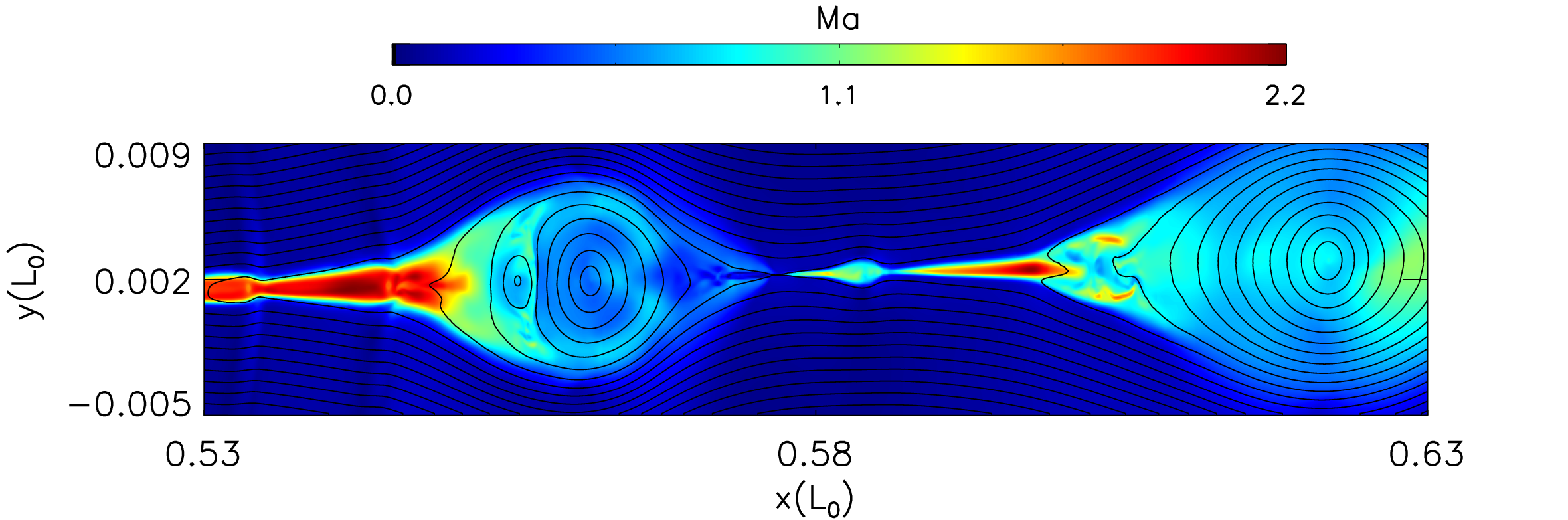}}
    \caption{ The distributions of current density, temperature and Mach number in the zoomed in small scale within $0.53L_0<x<0.63L_0$ and $-0.005L_0<y<0.009L_0$ in Case.~I at $t=1.592t_{A0I}$. The black arrows in the current density panel represent plasma velocities. }
\label{fig.4}
\end{figure*}

\begin{figure*}
 \centerline{\includegraphics[width=0.80\textwidth, clip=]{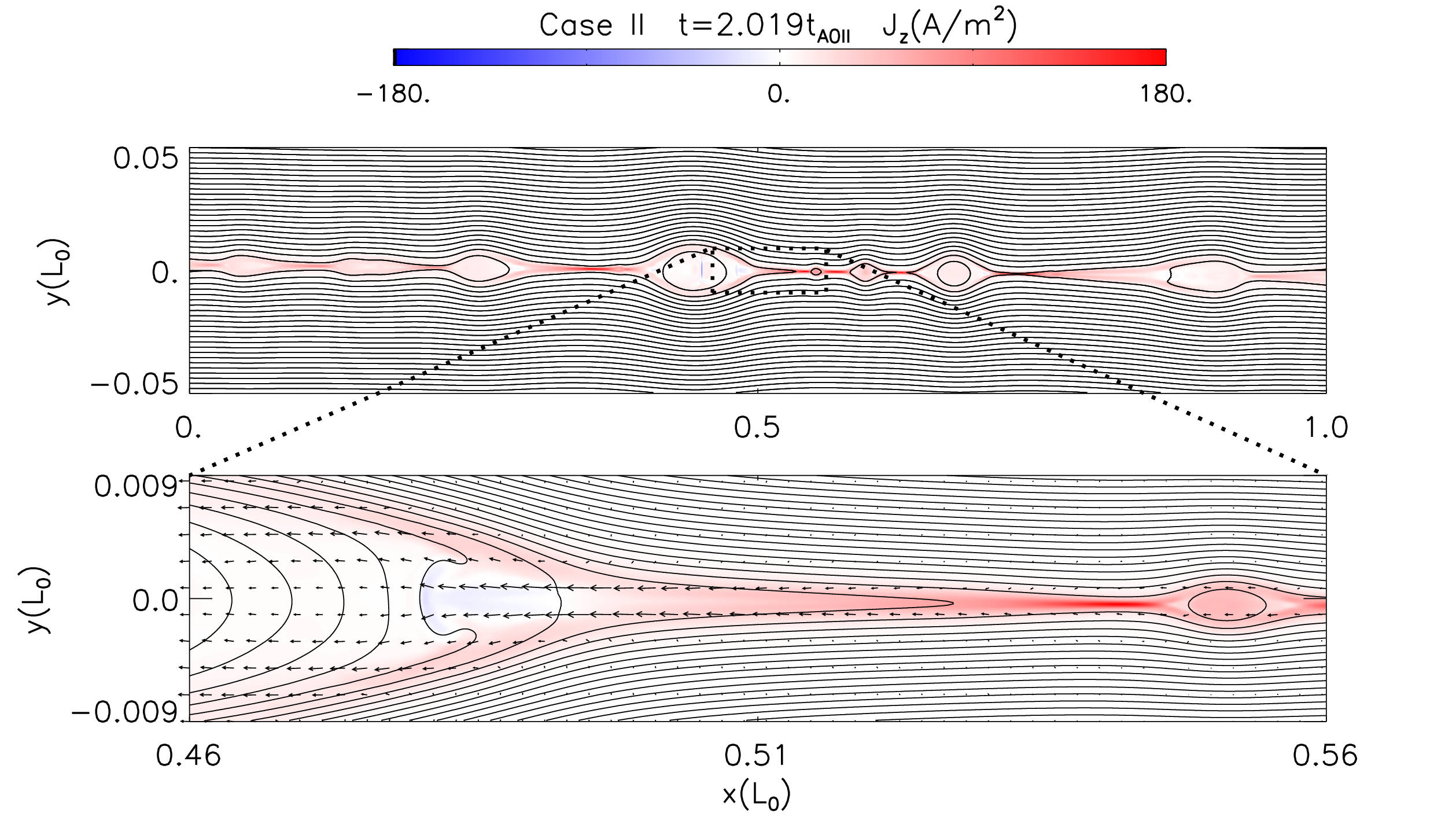}}
  \centerline{\includegraphics[width=0.80\textwidth, clip=]{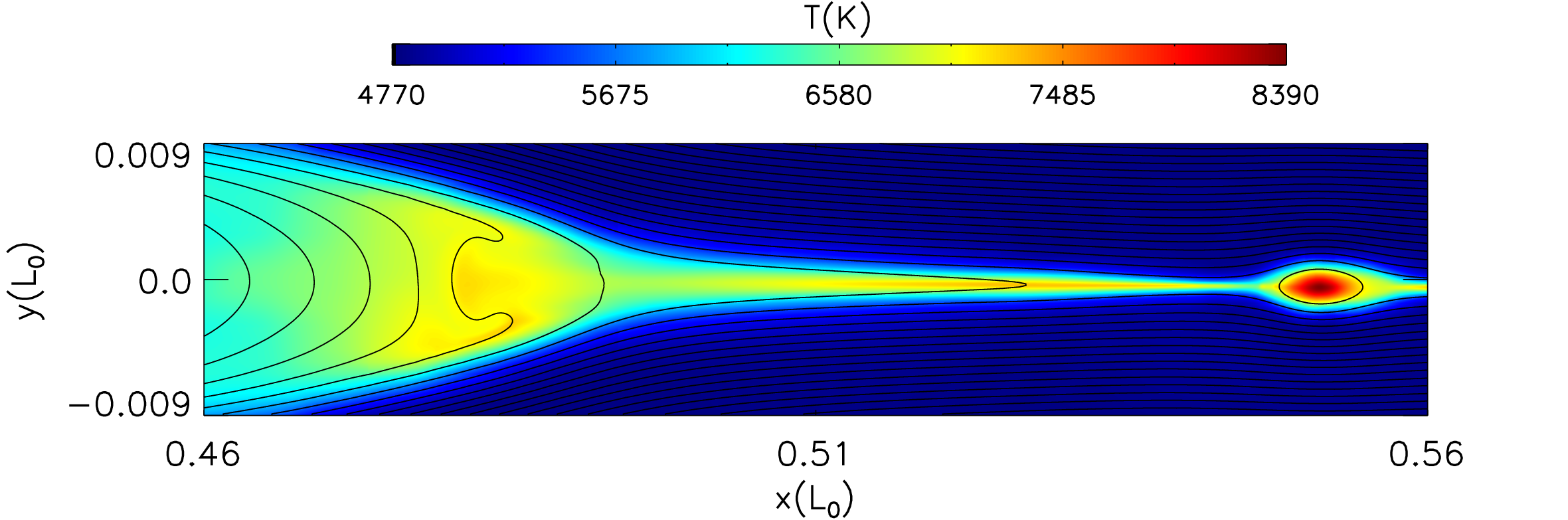}}
 \centerline{\includegraphics[width=0.80\textwidth, clip=]{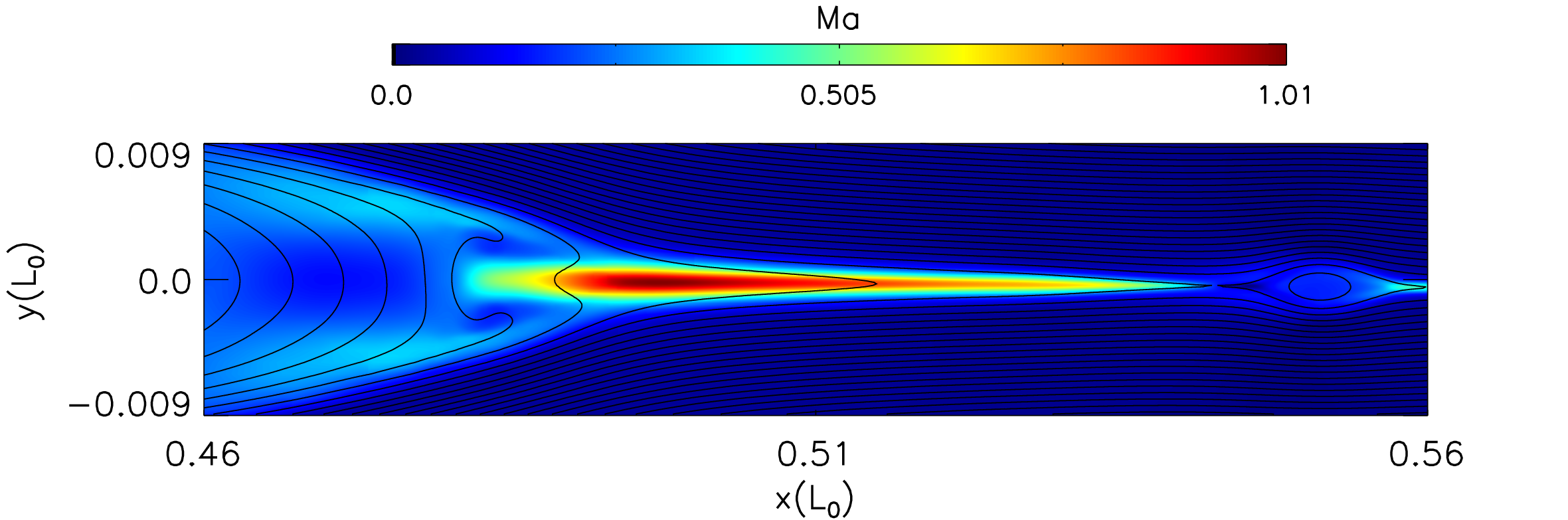}}
    \caption{The distributions of current density, temperature and Mach number in the zoomed in small scale within $0.46L_0<x<0.56L_0$ and $-0.009L_0<y<0.009L_0$ in Case.~II at $t=2.019t_{A0II}$. The black arrows in the current density panel also represent plasma velocities..}
\label{fig.5}
\end{figure*}

\begin{figure*}
 \centerline{\includegraphics[width=0.45\textwidth, clip=]{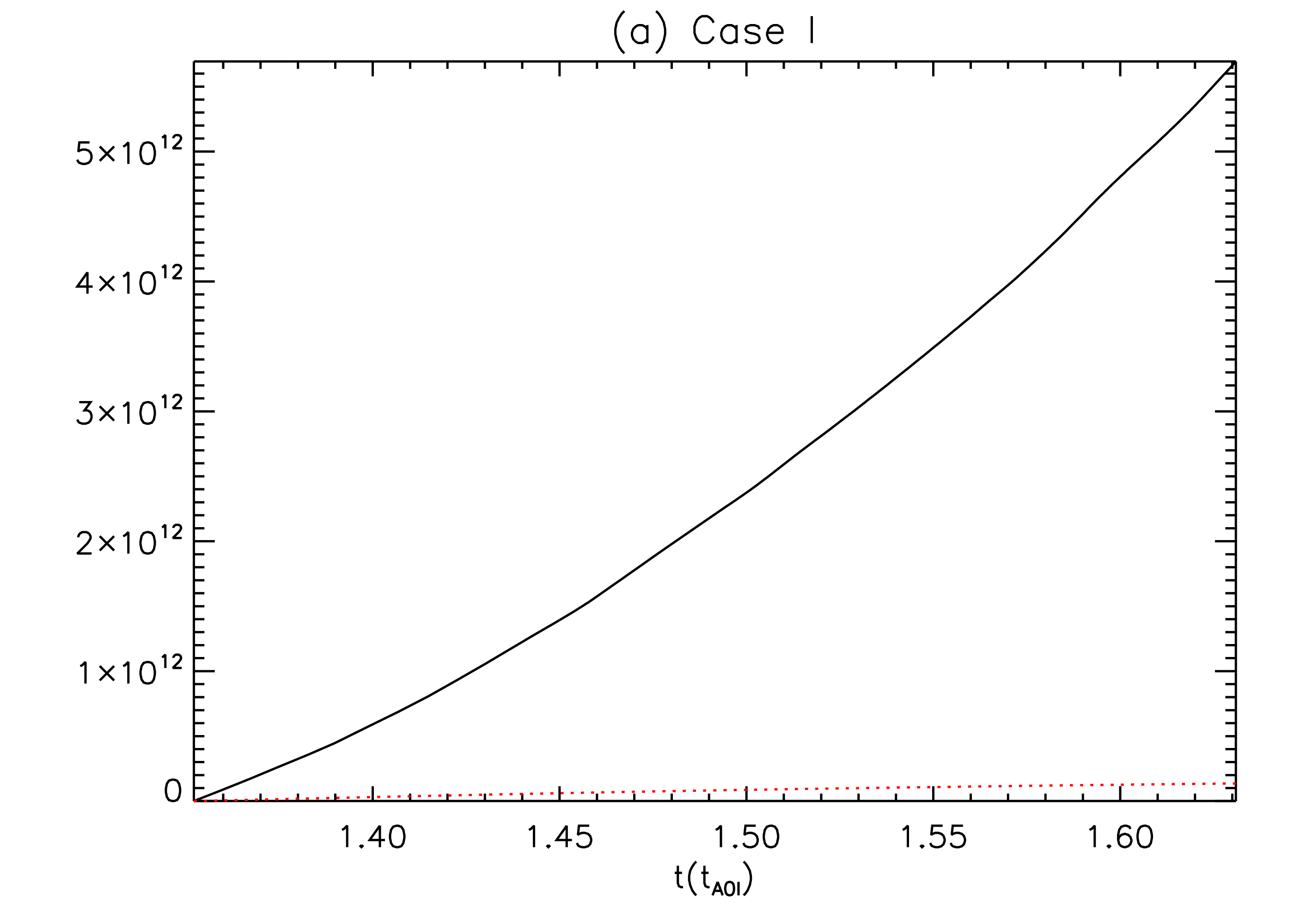}
                    \includegraphics[width=0.45\textwidth, clip=]{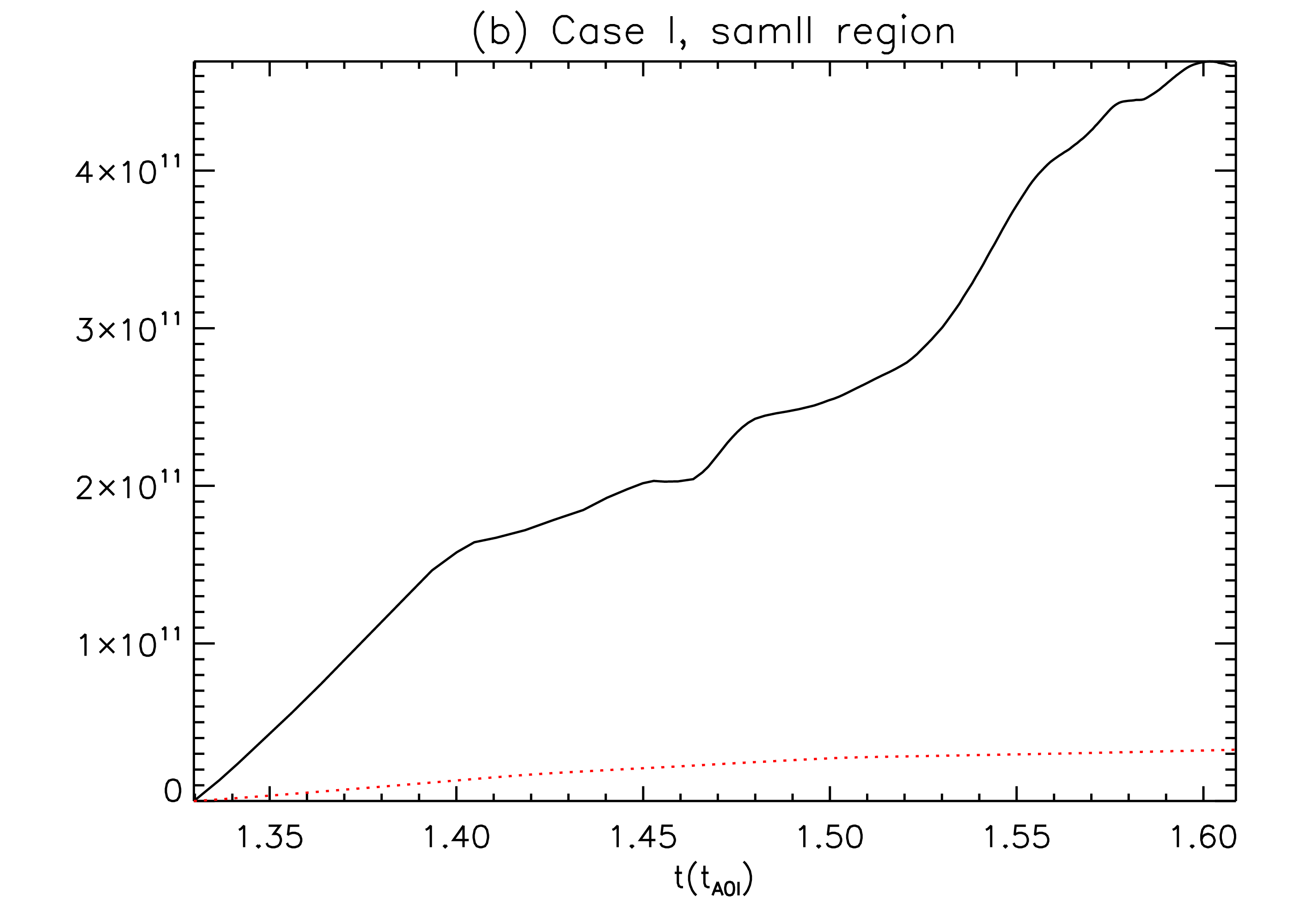}}
  \centerline{\includegraphics[width=0.45\textwidth, clip=]{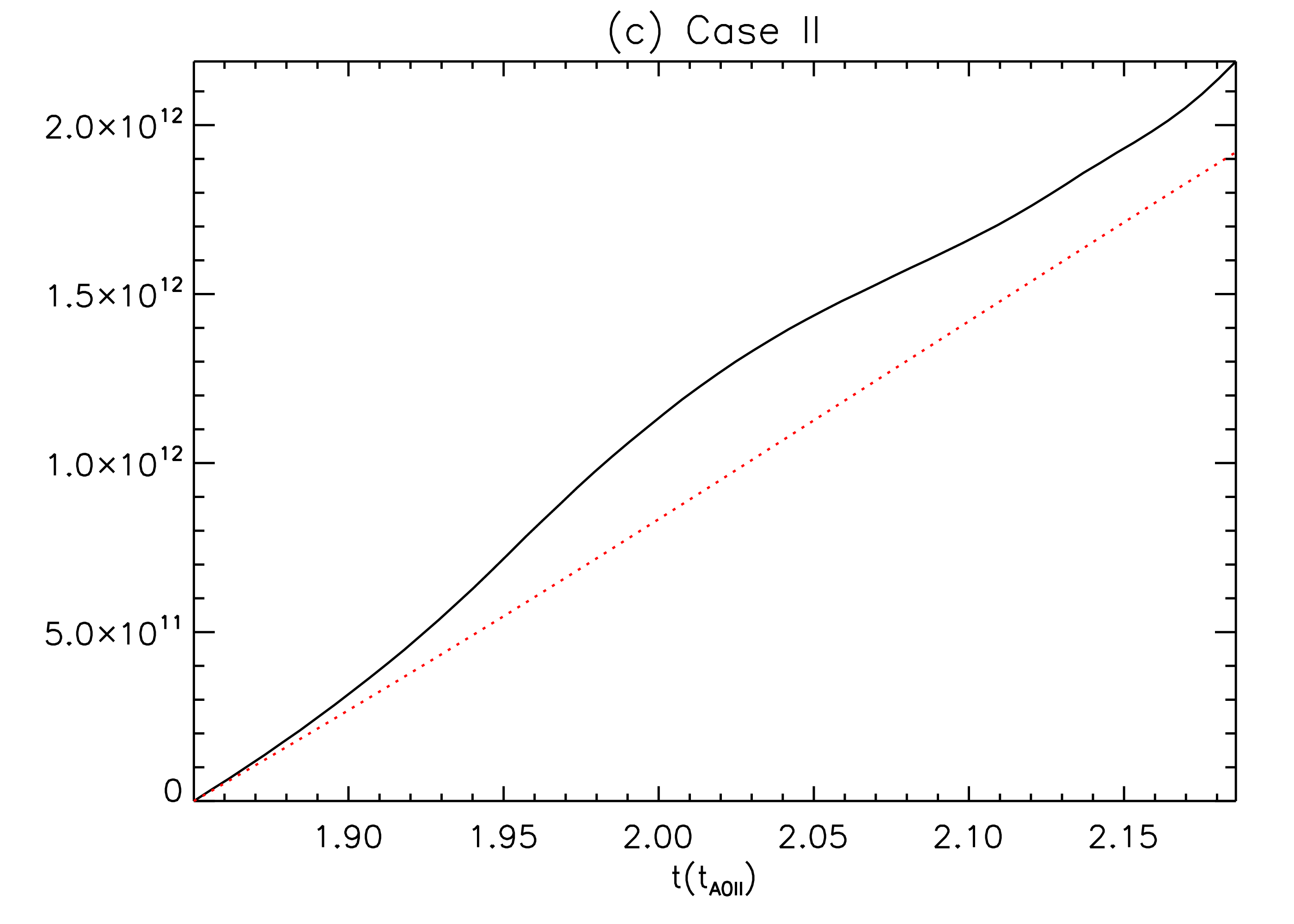}
                     \includegraphics[width=0.45\textwidth, clip=]{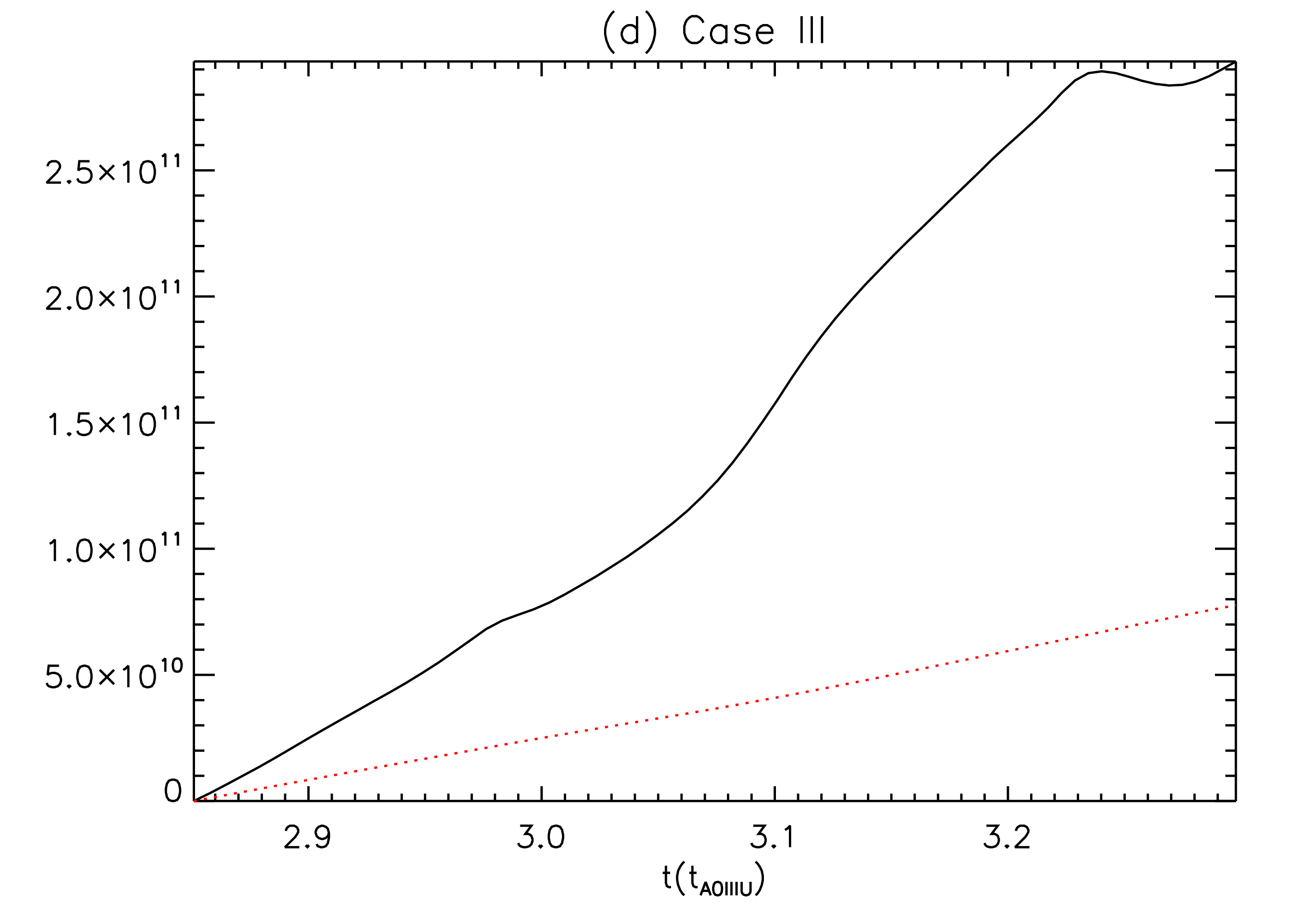}}
    \caption{The black solid lines represent the total generated thermal energy and  the red dotted lines represent the Joule heating, (a) calculated by using level 6 IDL data, in the domain $0.2L_0<x<0.8L_0$ and $-0.01L_0<y<0.01L_0$,  during the period $1.352t_{A0I}<t<1.631t_{A0I}$ in Case.~I; (b)calculated by using level 10 IDL data,  in the domain $0.55L_0<x<0.60L_0$ and $-0.002L_0<y<0.006L_0$ during the period $1.330t_{A0I}<t<1.609t_{A0I}$ in Case.~I; (c) calculated by using level 6 IDL data, in the domain $0.2L_0<x<0.8L_0$ and $-0.01L_0<y<0.01L_0$,  during the period $1.850t_{A0II}<t<2.186t_{A0II}$ in Case.~II.}
\label{fig.6}
\end{figure*}

\begin{figure*}
\centerline{\includegraphics[width=0.45\textwidth, clip=]{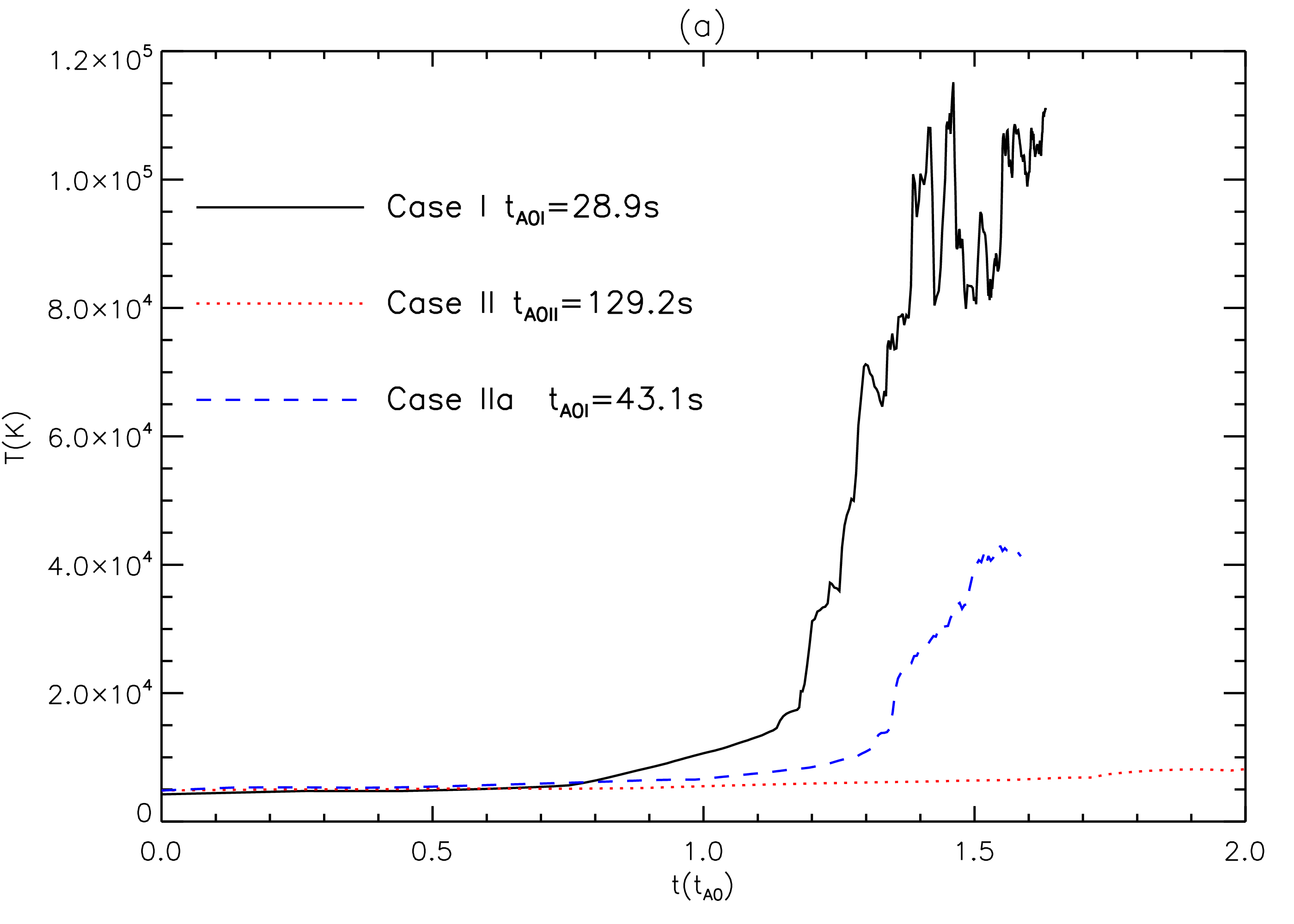}
                     \includegraphics[width=0.45\textwidth, clip=]{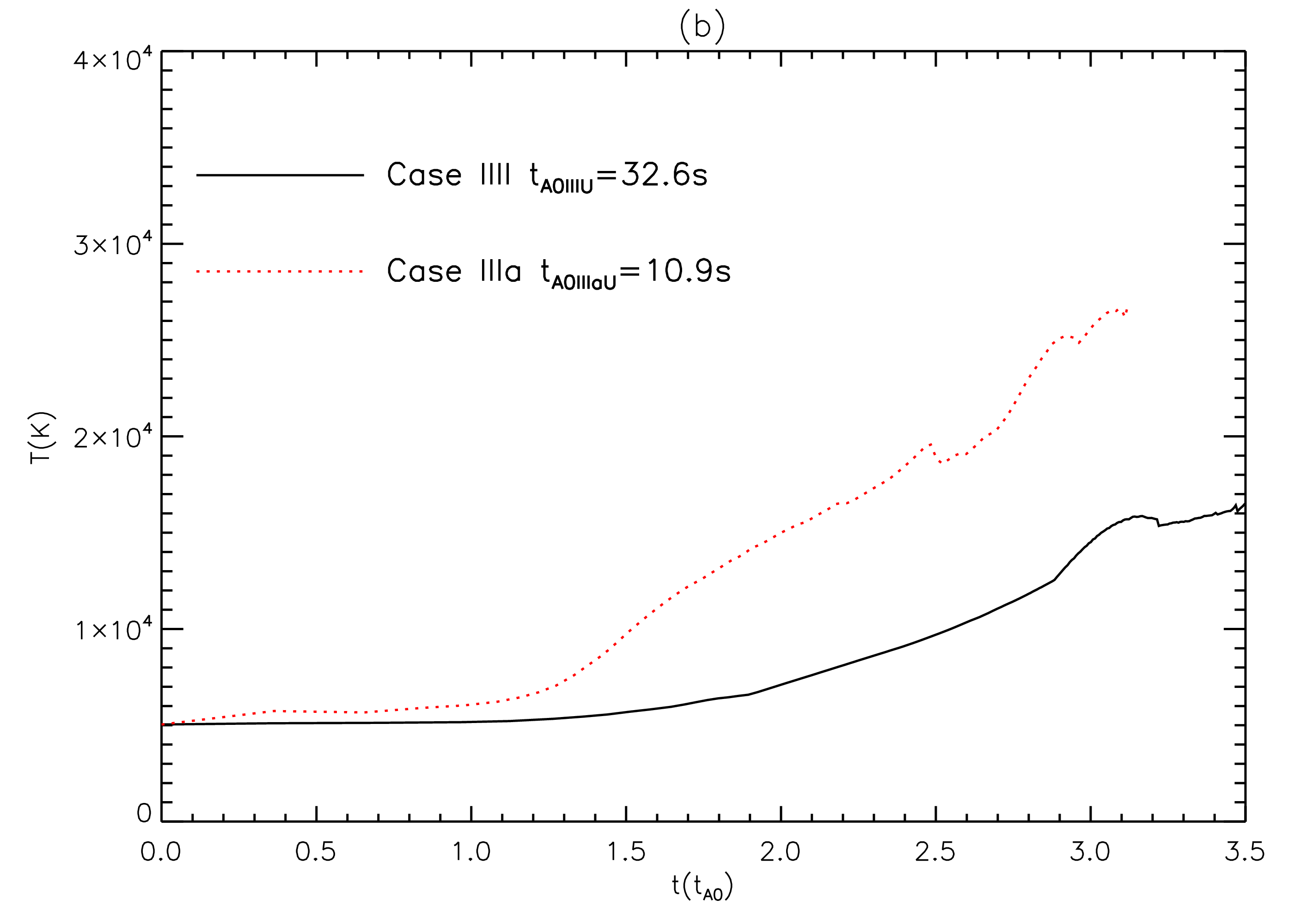}}
    \caption{The evolution of the maximum temperature in the current sheet region in (a) Case.~I, Case.~II and Case.~IIa; (b) Case.~III and Case.~IIIa. }
\label{fig.7}
\end{figure*}

\begin{figure*}
\centerline{\includegraphics[width=0.3\textwidth, clip=]{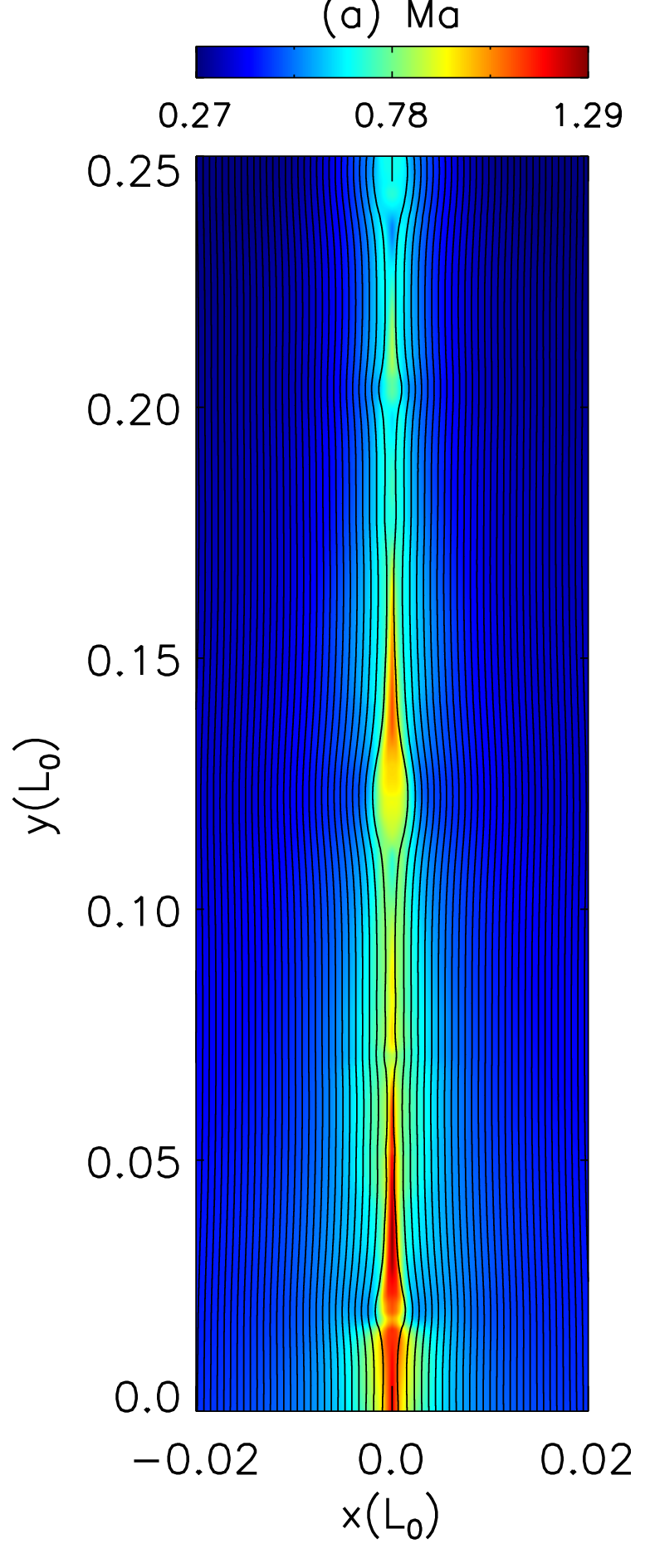}
                     \includegraphics[width=0.6\textwidth, clip=]{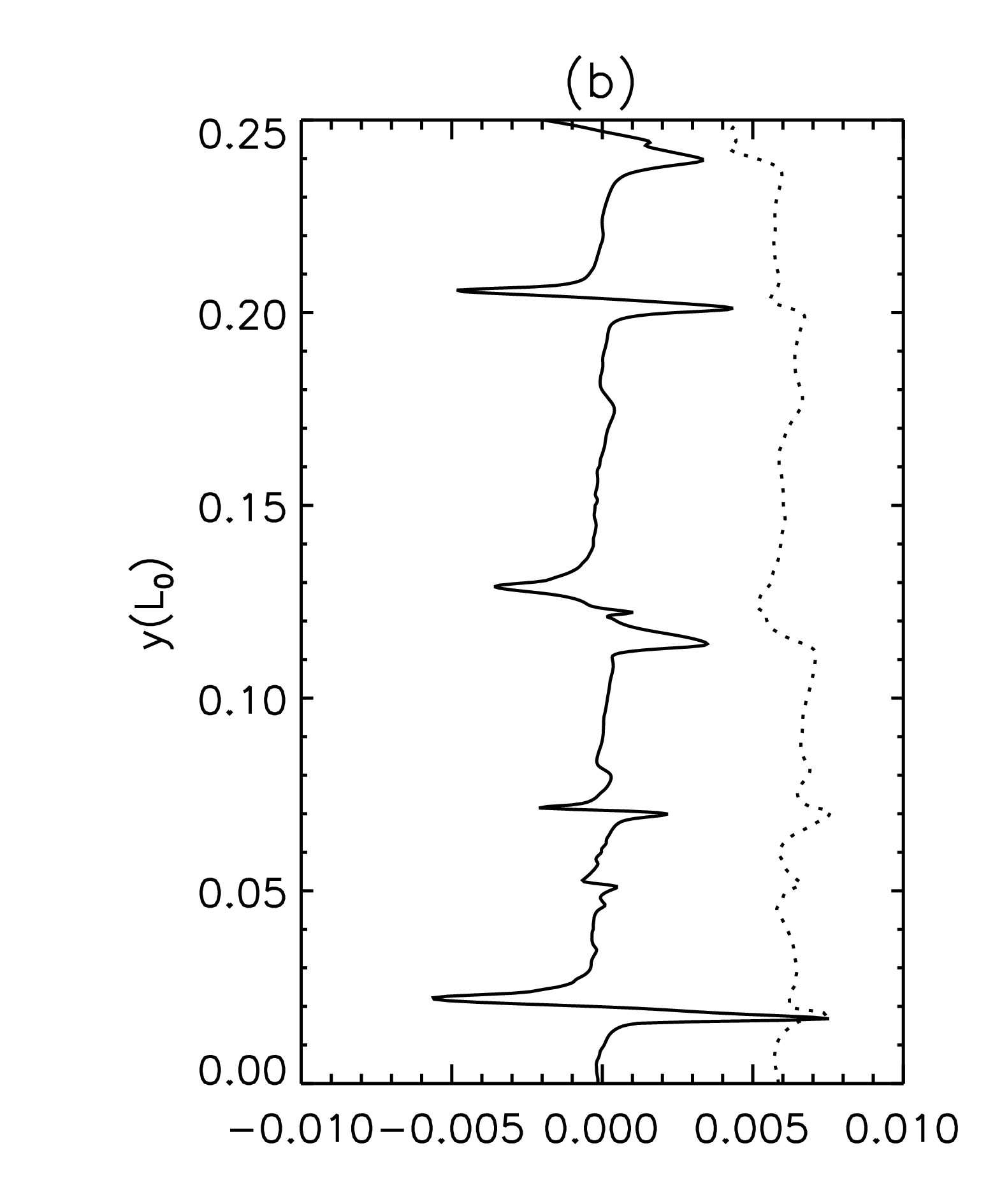}}
    \caption{(a)The distributions of the Mach number at $t=3.182t_{A0IIIU}$ in the domain $-0.02L_0<x<0.02L_0$ and $0<y<0.25L_0$; (b)The distributions of $B_x$ and $T/2\times10^6$ along y direction at $x=0$. The black solid line is $B_x$ and the dotted line is $T/2\times10^6$. }
\label{fig.8}
\end{figure*}

\end{document}